\documentclass[10pt, conference, compsocconf]{IEEEtran}
\usepackage{cite}
\usepackage{multirow}
\usepackage[pdftex]{graphicx}
\graphicspath{{../pdf/}{../jpeg/}}
\DeclareGraphicsExtensions{.pdf,.jpeg,.png}
\usepackage[cmex10]{amsmath}
\usepackage{amssymb}
\usepackage{algorithmic}
\usepackage{array}
\usepackage{mdwmath}
\usepackage{mdwtab}
\usepackage{eqparbox}
\usepackage{caption}
\usepackage[font=footnotesize]{subfig}
\usepackage{url}
\usepackage{booktabs}
\usepackage{color}

\newcommand\T{\rule{0pt}{2.9ex}}       % Top strut
\newcommand\B{\rule[-1.2ex]{0pt}{0pt}} % Bottom strut

\renewenvironment{thebibliography}[1]{
  \begin{oldthebibliography}{#1}
    \setlength{\itemsep}{0.1em}
    \setlength{\parskip}{0.13em}
}
{
  \end{oldthebibliography}
}
\begin{document}

\title{Temporal Graph Offset Reconstruction: Towards Temporally Robust Graph Representation Learning}

\author{
\IEEEauthorblockN{Stephen Bonner \IEEEauthorrefmark{2},
John Brennan \IEEEauthorrefmark{2},
Ibad Kureshi \IEEEauthorrefmark{5}, 
Georgios Theodoropoulos \IEEEauthorrefmark{4}, \\
Andrew Stephen McGough \IEEEauthorrefmark{3} and
Boguslaw Obara \IEEEauthorrefmark{2}}

\IEEEauthorblockA{\IEEEauthorrefmark{2}Department of Computer Science, Durham University, Durham, UK \\ \{s.a.r.bonner, j.d.brennan, ibad.kureshi, boguslaw.obara\}@durham.ac.uk} 
\IEEEauthorblockA{\IEEEauthorrefmark{3} School of Computing, Newcastle University, Newcastle, UK  \ stephen.mcgough@newcastle.ac.uk}
\IEEEauthorblockA{\IEEEauthorrefmark{4} School of Computer Science and Engineering, SUSTech, Shenzhen, China \ georgios@sustec.edu.cn }
\IEEEauthorblockA{\IEEEauthorrefmark{5} InlecomSystems, Brussels, Belgium  \ ibad.kureshi@inlecomsystems.com}}

% make the title area
\maketitle

\begin{abstract}

Graphs are a commonly used construct for representing relationships between elements in complex high dimensional datasets. Many real-world phenomenon are dynamic in nature, meaning that any graph used to represent them is inherently temporal. However, many of the machine learning models designed to capture knowledge about the structure of these graphs ignore this rich temporal information when creating representations of the graph. This results in models which do not perform well when used to make predictions about the future state of the graph -- especially when the delta between time stamps is not small. In this work, we explore a novel training procedure and an associated unsupervised model which creates graph representations optimised to predict the future state of the graph. We make use of graph convolutional neural networks to encode the graph into a latent representation, which we then use to train our temporal offset reconstruction method, inspired by auto-encoders, to predict a later time point -- multiple time steps into the future. Using our method, we demonstrate superior performance for the task of future link prediction compared with none-temporal state-of-the-art baselines. We show our approach to be capable of outperforming non-temporal baselines by 38\% on a real world dataset.
\end{abstract}


\IEEEpeerreviewmaketitle

%----------------------------------------------------------------------------------------
%	SECTION - Introduction
%----------------------------------------------------------------------------------------
\section{Introduction}
\label{sec:introduction}

Graphs have become a common metaphor in numerous scientific domains for representing high-dimensional complex data which is loosely topologically related, common examples being social media networks or citation networks. As such they provide a useful abstraction for how data is related. Graphs allow for complex analysis to be performed such as identifying the missing link within a graph (a person whom you might know or that paper you must read), however, to date almost all of the prediction work which has been performed on graphs has been focused on analysis in the topological domain as opposed to the temporal domain. This is strange as almost all graphs change with time (making new fiends or publishing new papers).

We formally define a graph $G = (V,E)$ as a finite set of vertices $V$, with a corresponding set of  of edges $E$. Elements of $E$ are unordered tuples $\{u,v\}$ where $u,v \in V$. Elements in $V$ and $E$ may have a label or some associated features, although these are not needed for this work. In order to perform analysis on graphs we need a mechanism which converts the formal graph representation into a format which is amenable for machine learning -- graph representation learning, sometimes referred to as graph embeddings.
%Unfortunately, the representation of a graph is not easily amenable to machine learning. Representations of a graph as unordered tuples are at too low a level to be of use for vast graphs whilst approaches such as adjacency matrix approaches lack a consistent `view' which can be used to compare between different graphs.

Recently, the field of graph representation learning has received a lot of interest as a way of analysing high-dimensional graphs via the use of machine learning. Graph representation learning, comprises a set of techniques which learn latent representations of a graph, which can then be used as input to machine learning models for downstream prediction tasks\cite{Grover2016}. The majority of graph representation learning techniques have focused upon learning vertex embeddings \cite{Goyal2017review} and reconstructing missing edges \cite{Grover2016}. As such, the goal of graph representation learning is to learn some function $f:V \rightarrow \mathbb{R}^d$ which maps from the set of vertices $V$ to a set of embeddings of the vertices, where $d$ is the required dimensionality. Resulting in $f$ being a matrix of dimensions $|V| $ by $ d$, i.e. an embedding of size $d$ for each vertex in the graph. However, the majority of graph representation learning approaches to date ignore the temporal aspect of dynamic graphs, resulting in models which have no concept of how the graph will change in the future. 

There are many cases in the real-world where a machine learning model will be trained using historical data and then used for inference at a later point in time. A primary example of this being the recommender systems literature, where graphs can be used to model the relationship between users and items \cite{berg2017graph}. Here a model is trained to recommend items to users based on historically collected data, and then used to make predictions about newly arriving data. However, the underlying graph structure is dynamic and could undergo large changes between time points, leading to the existing model suffering from degraded predictive performance, forcing the model to be retrained.

This paper introduces a new training procedure and associated model which can be used to create vertex level representations which encode information about how the graph will evolve into the future, leading to a model which is more robust to temporal change. To achieve this, we introduce the temporal offset reconstruction method to create graph representations which are explicitly designed to predict the next time point for a dynamic graph. We show that this offset method results in vertex representations which perform better when used to make predictions on later time points. Further, we make use of graph convolutional neural networks \cite{kipf2016semi}, combined with our temporal offset reconstruction method to show that graph convolutions can be used to capture the dynamics of an evolving graph dataset. To the best of our knowledge, this is the first time this has been done. In summary, we make the following contributions during this paper:

\begin{itemize}
  \item \emph{Temporal Offset Reconstruction} - We introduce the temporal offset reconstruction method to create vertex representations which contain temporal, in addition to structural, information about the vertex. Achieved by explicitly training a model to predict the future rather than the current state of the graph.
  
  \item \emph{Graph Convolution-based Model} - We explore the use of unsupervised graph convolutional networks to capture graph dynamics. 
  
  \item \emph{Detailed Experimentation} - We perform detailed experimentation on the task of future link prediction using benchmark graph datasets, containing synthetic and empirical evolutionary graphs. 
\end{itemize}

To aid the reproducibility of our work, we open-source all of our PyTorch \cite{paszke2017automatic} based source-code\footnote{\url{https://github.com/sbonner0/temporal-offset-reconstruction}} and experimentation scripts, as well as presenting results upon benchmark datasets. 

The remainder of this paper is structured as follows: In Section \ref{sec:litreview} we review the relevant literature, Section \ref{sec:method} we discuss our methodology and introduce our temporal offset approach, Section \ref{sec:experiment_setup} details the experimental setup,  Section \ref{sec:results} presents the results and Section \ref{sec:conclusion} draws conclusions.

\section{Related Works}
\label{sec:litreview}
% Good review of graph conv nets here - https://arxiv.org/pdf/1806.00088.pdf
\subsection{Graph Representation Learning}

% The field of graph representation learning has attracted a large amount of interest from the research community, and a large number of competing approaches have emerged. For a complete overview of the field, readers are refereed to one of the recent surveys \cite{hamilton2017representation, goyal2018graph, bonner2017evaluating}, However, we will review the literature directly relevant to our work here. The family of graph embedding approaches can be categorised in a variety of ways - examples of possible categorisations include how the graph is represented for the input (adjacency matrix / walks), the type of learning performed (unsupervised, supervised), the type of machine learning model used (matrix factorisation/ skip-gram / deep learning) and even the space in which the embeddings reside (euclidean / hyperbolic). 

Traditionally graph representations were created via techniques based on matrix factorization, where a mapping to a lower dimensional space is found such that pair-wise relationships in the original graph are preserved. Examples of such approaches include Laplican eigenmaps \cite{belkin2002laplacian}, Graph Factorization \cite{ahmed2013distributed}, GraGrep \cite{cao2015grarep} and HOPE \cite{ou2016asymmetric}. More recently, models originally designed for Natural Language Processing (NLP) have been adapted to create graph embeddings. Such approaches use random walks to create `sentences' which can be used as input to language inspired models such as Word2Vec \cite{mikolov2013distributed}. NLP inspired graph embedding approaches include DeepWalk \cite{Perozzi2014}, Node2Vec \cite{Grover2016} and Hyperbolic embeddings \cite{Chamberlain2017}.

Recently graph specific neural network based models have been created which are inspired by Convolutional Neural Networks (CNNs) from the field of computer vision. Such approaches attempt to create a differential model for learning directly from graph structures. Many Graph CNN approaches operate in the spectral domain of the graph, using eigenvectors derived from the Laplacian matrix of a graph \cite{kipf2016semi}. Early approaches to define convolution operators on graphs often had large memory and computation complexities and were thus unsuited to many real world graphs~\cite{bruna2013spectral}.  

Later more efficient spectral methods were proposed which reduced the complexity of the filtering operations whilst still operating on the entire adjacency matrix \cite{defferrard2016convolutional, kipf2016semi}. The Graph Convolutional Network (GCN) approach has proven to be particularly effective \cite{kipf2016semi}. The GCN approaches uses a layer-wise propagation rule to aggregate information from a vertices 1-hop neighbourhood to create it's representation, this layer-wise rule can be stacked k times to aggregate information from k-hops away from a given vertex. The requirement to have the whole adjacency matrix available in memory means that the GCN approach struggles to scale to massive graphs. To tackle this problem, Graph-Sage \cite{hamilton2017inductive} learns to aggregate features from a fixed size \emph{sample} of a vertices neighbourhood and as such can be applied to new vertices which have joined the graph. However, the approach mandates that all vertices in the graph have features available and the performance can vary depending upon the neighbourhood sampling strategy~\cite{chen2018fastgcn}. 

Thus far, all the graph specific models discussed have been supervised approaches, requiring the graph's to have labels. There have been a modest selection of unsupervised graph specific neural models, many are based on auto-encoders - a type of neural network whose task is to reconstruct the input data after being projected into a lower-dimension \cite{baldi2012autoencoders}. Structural Deep Network Embedding (SDNE) uses an auto-encoder to reconstruct each row in a graph's adjacency matrix \cite{wang2016structural}. A more recent auto-encoder employs a generative model to adversarially regularise the embeddings to help improve performance \cite{yu2018learning}. Work has also explored the use of GCNs as the basis of a convolutional auto-encoder model \cite{kipf2016variational}, producing state-of-the-art results for link-prediction in citation graphs. 

\subsection{Temporal Embeddings}

% All of the embedding approaches discussed so far have considered stationary non-evolving graphs. This section will review the literature regarding attempts to create temporal embeddings. As many graph embedding techniques are taken from NLP, we briefly review the approaches here which consider the evolution of language, before looking at graph specific approaches. 

\subsubsection{Natural Language Temporal Embeddings}

Some approaches from temporal language based models are created to model how word use evolves over time. For example, work has been performed to use cosine similarity to automatically measure how a word changes, relative to it's neighbours, over time and identify anomalies \cite{Kim2014}. To overcome this non-convex problem, work has been performed in creating diachronic word embedding by aligning different embedding snapshots using orthogonal Procrustes, making the learning not end to end \cite{Hamilton2016}. In later work, a dynamic Word2Vec model for word embedding is created which attempts to solve the non-convex problem common to embeddings via Bayesian variational black-box inference \cite{Bamler2017}. The approach creates embeddings which change smoothly over time and are better able to predict the change in context for a given word than previous methods.

\subsubsection{Graph Specific Temporal Embeddings}

To date, there have been few attempts to consider the temporal change of a graph when creating its embedding. However the existing approaches can broadly be split into two categories: Temporal Walk and Adjacency Matrix Factorisation based.

\emph{Temporal Walk-Based Approaches - }

Many of the temporal graph embedding approaches which exist are based on data created via temporal walks, which are random walks over dynamic graphs. Perhaps the first such approach is that of STWalk \cite{pandhre2018stwalk}. In this work, the authors aim to learn \emph{node trajectories} via the use of random walks which learn representations that considers all the previous time-steps of a temporal graph. In the best performing approach presented, the authors learn two representations for a given vertex simultaneously which are concatenated to create the final embedding. The first representation is a normal DeepWalk embedding designed to capture the spatial information for a vertex. Wile the second representation is learned across a specially constructed graph structure, where each vertex is connected to it's 1-hop neighbourhood from each previous time-step. However the approach is not end-to-end and requires the user to manually chose how many time steps to consider. 

% NetWalk: A Flexible Deep Embedding Approach for Anomaly Detection in Dynamic Networks -----------------
Yu et al.~\cite{Yu2018}, propose NetWalk, a vertex-level dynamic graph embedding model using random walks designed to facilitate anomaly detection in streaming graphs. The approach captures a collection of short random walks from the graph which are then passed into an auto-encoder based model to create the vertex representations. In addition to the usual reconstruction based loss term, an additional term is added to minimise the pair-wise distance between the representations of vertices occurring within the same walk. To apply this approach to the domain of streaming graphs, where changes to graph are being made online, the approach maintains for each vertex a list of neighbouring vertices which is updated as the graph changes. If changes in a vertices neighbour list occur, new random walks will be generated and the representations updated. The final anomaly detection is performed via a dynamic clustering model on the vertex representations. However, the created embeddings are not capable of capturing temporal dynamics or reliably predicting the future state of a graph.

% Continuous-Time Dynamic Network Embeddings -----------------
Nguyen et al.~\cite{nguyen2018continuous}, propose a model to incorporate temporal information when creating graph embeddings via random walks by capturing individual changes (edge addition/deletion for example) within a graph. The authors propose a temporal random walk to create the input data, however their approach creates more complex and rich temporal walks via a biasing process. The approach can be used to add temporal information into any embedding model which relies on random walks as input data, however the paper explicitly explores a model based on the Skip-Gram architecture and shows the predictive performance increases over non-temporal baselines.

\emph{Adjacency Matrix Factorisation Approaches - }

% DynGEM: Deep Embedding Method for Dynamic Graphs -----------------
Goyal et al.~\cite{Goyal2017}, propose a model for creating dynamic graph embeddings, entitled DynGEM. In this approach they extend the auto-encoder graph embedding model of SDNE~\cite{wang2016structural} to consider dynamic graphs. To do this, they use a method similar to Net2net~\cite{Chen2015}, which is designed to transfer the learned knowledge from one neural network to a second model. This technique allows them to add more neurons to the auto-encoder, appropriate to the increasing graph size, via a heuristic approach entitled PropSize. The use of the Net2net technique means that the model can be expanded while ensuring the learned function is approximately preserved. The process for training the model is as follows, at the fist time snapshot of the graph, a complete graph auto-encoder is trained as described in the SDNE paper \cite{wang2016structural}. For the next time step of the model, a new auto encoder is trained reusing the weights from the previous step, with any new neurons being added according to the PropSize heuristic. However the approach does not explicitly predict the future state of the graph, rather, it transfers knowledge from the previous timestamp to help the current auto-encoder better reconstruct the current time-step. 

Additionally there have been attempts to incorporate temporal aspects into GCNs \cite{manessi2017dynamic,seo2016structured}, however both approaches focus upon supervised learning and do not explicitly use the models to predict the future graph state. 

%----------------------------------------------------------------------------------------
%	SECTION - Method
%----------------------------------------------------------------------------------------
\section{Method}
\label{sec:method}

\subsection{Motivation}
\label{sec:motivation}

Many empirical graph datasets comprise a stream of continually evolving changes, as connections within the graph are formed and broken. One can consider the graph then as a series of snapshots, with each snapshot contains the graph as it is at that point in time. More formally, we can redefine a graph $G$ to be a temporal graph $G^\prime = \{G_0, G_1,...,G_t\}$, where each graph snapshot $G_{i}$ $(I \in [0,t])$  contains a corresponding vertex set $V_{i}$ and edge set $E_{i}$.

In many real-world use cases of machine learning, a model is trained on historical data and then used to make predictions about new events at a future point in time. An example of where this practice is common is in the recommender systems industry where recent state-of-the-art systems, for recommending items to users, are based on graph convolutions \cite{berg2017graph, ying2018graph}. However, to date, the majority of models for creating graph representations do not consider how the graph evolves over time. This could potentially result in models which have good initial predictive capability, but whose performance will degrade as the graph continues to change over time. 

\subsection{Temporal Graph Offset Reconstruction}
\label{sec:tor}

We propose to tackle the challenging problem of creating temporal robust graph embeddings by training a model to explicitly recreate a future time step of the graph. We name this approach temporal offset reconstruction as the input and target graph are offset by at least one time point. More concretely, a graph $G_i$ is used as input to model $\theta(G_i)$ which learns a representation for each vertex in $G_i$ such that it's output can accurately predict the graph $G_{i+\delta}$. Ideally, we want to create a model $\theta(G_i)$ which can perform this temporal offset reconstruction using the graphs $G_i$ and $G_{i+\delta}$ alone,  $G_{i+\delta} = \theta(G_i)$, requiring no pre-processing steps which could affect the models performance (e.g. random walk procedures), no pre-computed vertex features and no labels required or used.

In the remainder of this section, we will detail the graph convolutions used to create the vertex representations, the models we explore to perform the temporal offset reconstruction and the training procedure. 

\subsection{Graph Convolutions}
\label{sec:gcn}

To perform the graph encoding to create the initial vertex representations, we propose to utilise the spectral graph convolution networks introduced by Kipf \cite{kipf2016semi}. GCNs are fully differential end-to-end models for learning from graphs, meaning that no pre-processing step is required. 

One can consider a GCN to be a differentiable function for aggregating information from a vertices immediate neighbourhood \cite{chen2018fastgcn,hamilton2017inductive}. The GCN framework introduced by Kipf produces a representation $Z$ for each vertex in a graph, provided in the form of it's adjacency matrix\footnote{The adjacency matrix $A = (a _i,_j)$ for a graph is symmetric matrix of size $|V|$ by $|V|$, where $(a _i,_j)$ is 1 if an edge is present and 0 otherwise.}  $A$ with a set of vertex level features $X$, taking the form of $Z = GCN(A, X)$. When no vertex features are present the identity matrix $I$ of $A$ can be used such that $X = I$. As with any deep neural network, a GCN can contain many layers which aggregate the data. The operation performed at each layer by the GCN is as follows \cite{kipf2016semi}:
\begin{equation}
    \label{eq:GCN}
    GCN^{(l)}(H^{(l)}, \hat A) = \sigma (\hat AH^{(l)}W^{(l)}) \, ,
\end{equation}
where $l$ is the current layer number, $W^{(l)}$ is the weight matrix of that layer, $H$ are the features computed at the previous layer or is equal to $X$ at $l=0$, and $\sigma$ is a non-linear activation function like the rectified linear unit (ReLU). To help with feature scaling, $\hat A$ is the matrix $A$ normalised by it's degree matrix $D$ and it's identity matrix $I$ such that $\hat A = ({D}^{-\frac{1}{2}} (A+I) {D}^{-\frac{1}{2}})$ \cite{kipf2016semi}. 

Essentially, we can consider this GCN function to be aggregating a weighted average of the neighbourhood features for each vertex in the graph. Stacking multiple GCN layers has the effect of increasing the number of hops from which a vertex-level representation can aggregate information - for example, a three layer GCN will aggregate information from three-hops within the graph to create each representation. 

The original methods presented in the literature required GCN based models to be trained via supervised learning, where the final vertex representation is tuned via provided labels for a specific task -- classification as a common example \cite{kipf2016semi,hamilton2017inductive}. This is a key difference between GCNs and other graph embedding approaches, as these commonly require no labels and thus are applicable on a broader selection of graphs. Recently, extensions to the GCN framework have been made which allows for convolutional auto-encoders for graph datasets \cite{kipf2016variational}. Auto-encoders are a type of un-supervised neural network model which attempt to compress input data to a low-dimensional space, and then reconstruct the original data directly from the learned representation. We will explore similar concepts but using them for temporal reconstruction in the following section.

\subsection{Model Overview}
\label{sec:model_overview}

For creating our temporally offset graph embeddings, we will explore the use of both non-probabilistic and variational encoder models. These are related to the convolutional graph auto-encoders of Kipf \cite{kipf2016variational}, however, we are exploring the creation of two  models explicitly to reconstruct a future state of the graph, rather than just to capture the current graph. Below we detail the specifics of both the non-probabilistic (TO-GAE) and variational (TO-GVAE) models used for temporal offset reconstruction.

\textbf{TO-GAE:} TO-GAE is the non-probabilistic interpretation of the temporally offset graph auto-encoder concept, where the goal is to learn a low-dimensional representation of $A_i$ from $G_i$, via an encoding from a GCN $Z_i = GCN(A_i, X_i)$, such that it can be used to predict accurately the structure of some future time step $\delta$ of the graph via a product between $Z_i$ and it's transpose passed through a logistic sigmoid unit $\sigma$:

\begin{equation}
    \label{eq:GAE}
    A_{i+\delta} = \sigma(Z_i Z_i^\mathsf{T}).
\end{equation}

For all the work presented in the paper, the GCN model used to learn $Z_i$ is a two layer model. 

\textbf{TO-GVAE:} TO-GVAE is a variational interpretation of the temporally offset graph auto-encoder concept. Again the goal is to learn a vertex level representation for future graph reconstruction but using ideas from Bayesian learning \cite{kingma2013auto}. This variational method differs from the non-probabilistic version outlined above as instead of directly learning the mapping $Z$, we instead learn a distribution from which $Z$ is sampled. Using a variational approach to create the latent space has been shown to create more robust and meaningful embeddings, resulting in better performing models \cite{kingma2013auto, kipf2016variational}.

As TO-GVAE is a Bayesian style model, we must define a model with which to perform inference. This again makes use of the GCN layers outlined in Section \ref{sec:gcn} to learn a mean $\mu$ and a variance $\gamma$ vector used to parametrise the Gaussian distribution $\mathcal{N}$ from which $Z_i$ is finally sampled: 

\begin{equation}
    \label{eq:GVAE-inf}
    q(Z_i|X_i, A_i) = \prod_{v=1}^{|V|} \mathcal{N}(z_v | \mu_v, \, diag(\gamma^{2}_v))
\end{equation}

where $q$ is our approximation of the of the true distribution we are interesting in capturing --  $p(A_{i+\delta}|Z_i)$. Here both $\mu$ and $\gamma$ are separate individual GCN layers $GCN^{(\mu)}$ and $GCN^{(\gamma)}$ respectively. These layers however take their input from a common first layer $GCN^{(0)}$.

Once the inference model specified in Equation \ref{eq:GVAE-inf} has created the various required parameters, a generative model is created to predict the next time step in the graph. The generative model we use is again based on the inner-product between the latent representations, given as follows:

\begin{equation}
  \label{eq:GVAE-gen}
  p(A_{i+\delta}|Z_i) = \prod_{x=1}^{|V|} \prod_{y=1}^{|V|} p(A_{xy}=1|\sigma(z_x z_y^\mathsf{T})),
\end{equation}
where $A_{xy}$ are the elements from $A_i$ and $z$ are the rows for each vertex taken from $Z_i$.

To train the model, common for variational methods \cite{kingma2013auto, kipf2016variational}, we directly optimise the lower bound $L$ with regards to the model parameters: 

\begin{equation}
  \begin{gathered}
    \label{eq:GVAE-learn}
    L = \mathbb{E}_{q(Z_i|X_i, A_i)} \Big[ \text{log} p(A_{i+\delta}|Z_i) \Big] - \\ 
    KL(q(Z_i| A_i, X_i) || p(Z_i)),
  \end{gathered}
\end{equation}
%TODO what is \mathbb{E}
where KL is the Kullback-Leibler distance between p and q. For this model we use a Gaussian prior as the distribution for $p(Z_i)$. 

\subsection{Model Parameters and Training Procedure}
\label{sec:model_parm}

After initial experimentation we found two layers of convolution to give the optimal performance for both models, with the first layer comprising 32 filters, whilst the second having 16 filters. For training both the models we use full-batch gradient descent via the ADAM algorithm with a learning rate of 0.001 for a total of 50 epochs. We also found the use of an additional term in the loss functions which penalises the model parameters for getting too large via L2 to help model performance. All of our models, as well as the comparative baselines, have been implemented in the PyTorch library \cite{paszke2017automatic}. 

% http://kvfrans.com/variational-autoencoders-explained/ Good explaination of VAEs This constraint forces the encoder to be very efficient, creating information-rich latent variables. This improves generalization, so latent variables that we either randomly generated, or we got from encoding non-training images, will produce a nicer result when decoded.

%----------------------------------------------------------------------------------------
%	SECTION - Experimental Setup
%----------------------------------------------------------------------------------------
\section{Experimental Setup}
\label{sec:experiment_setup}

\subsection{Evaluation Overview and Methodology}
\label{sec:testingmeth}

As the primary goal of our approach is to create vertex representations which are better able to encode information about how the graph will evolve into the future, we will be using the important task of link prediction as our primary task. Formally the task of link prediction in the context of machine learning can be defined as follows: given a subset of edges $E_{train} \subset E$ from graph $G=(V,E)$, learn a model which can accurately predict the remaining edges $E_{test} = E - E_{train}$ \cite{martinez2017survey}. Many recent methods attempt to solve this problem via vertex embedding similarity -- i.e. vertices with more similar embeddings, according to some metric, are more likely to be connected via an edge \cite{Grover2016, Perozzi2014, kipf2016variational}. 

During the evaluation, we will be investigating two ways a model trained on temporal graph data could be used for inference:

  \begin{itemize}
    \item \emph{Evolution Pattern Prediction} -- For this inference approach, the original input graph $G_0$ is kept constant, whilst the rest of the time series $G_1,...,G_t$ is used as the targets for prediction to measure the models ability to predict the future graph changes.
     
    \item \emph{Future Link Prediction} -- For this inference approach, the trained model is kept constant whilst each graph snapshot $G_i$ is passed in. The model is then evaluated by making predictions on hold out edges from the same snapshot to test how well the model can predict edges in future snapshots of the graph.

  \end{itemize}

Graph edges are predicted as follows: Given the learned vertex embeddings, the adjacency matrix is reconstructed via a dot-product of the embedding matrix $A^{\prime} = \sigma(ZZ^\mathsf{T})$. This reconstructed adjacency matrix is compared with the true graph to assess how well the embedding is able to reconstruct the future graph. For the future link prediction task, edges from the graph are randomly removed before the graph is fed into the model. The embeddings are then used to predict the hold-out set of edges. 

\subsection{Performance Metrics}
\label{sec:metrics}

As one can consider the task of link prediction to be that of a binary classification problem (an edge can only be present or not), we make use of two standard binary classification metrics:

  \begin{itemize}
    \item \emph{Area Under the Receiver Operating Characteristic Curve (AUC)} -- The ratio between the True Positive Rate (TPR) and False Positive Rate (FPR) measured at various classification thresholds. 
    \item \emph{Mean Average Precision (AP)} -- The mean average precision across the set of test edges: $ Precision (AP) = \frac{ TP } {TP+FP },$ where $TP$ denotes the number of true positives the model predicts, and $FP$ denotes the number of false positives.
  \end{itemize}
In both cases large values are better.
% PUT HERE FOR LAYOUT ISSUES
  \begin{figure*} 
    \centering
  \subfloat[AUC score]{%
      \includegraphics[width=0.25\linewidth]{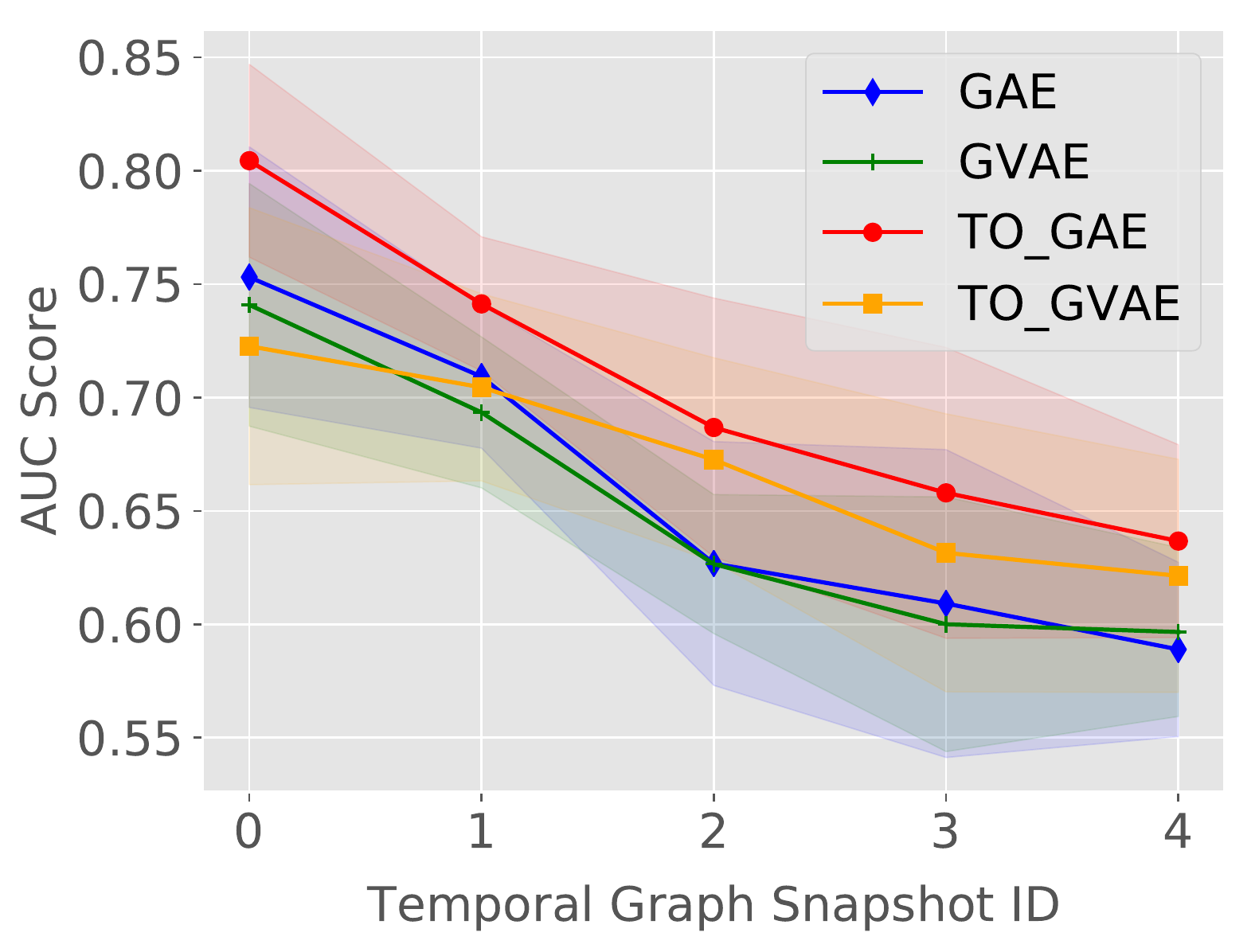}}
    \label{PRmiCA1}\hfill
  \subfloat[AUC score on new edges]{%
      \includegraphics[width=0.25\linewidth]{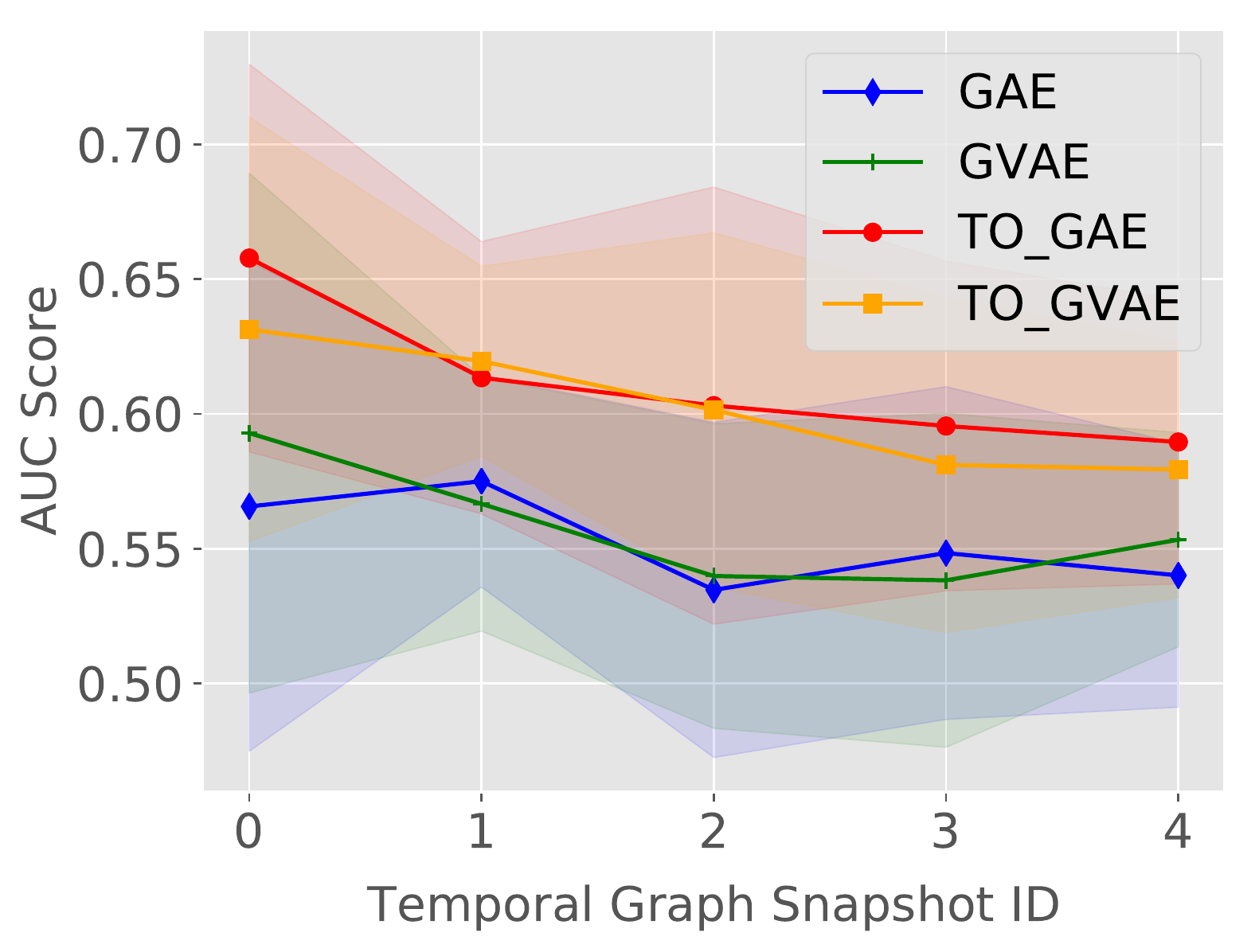}}
    \label{PRmiFB1}\hfill
  \subfloat[AP score]{%
      \includegraphics[width=0.25\linewidth]{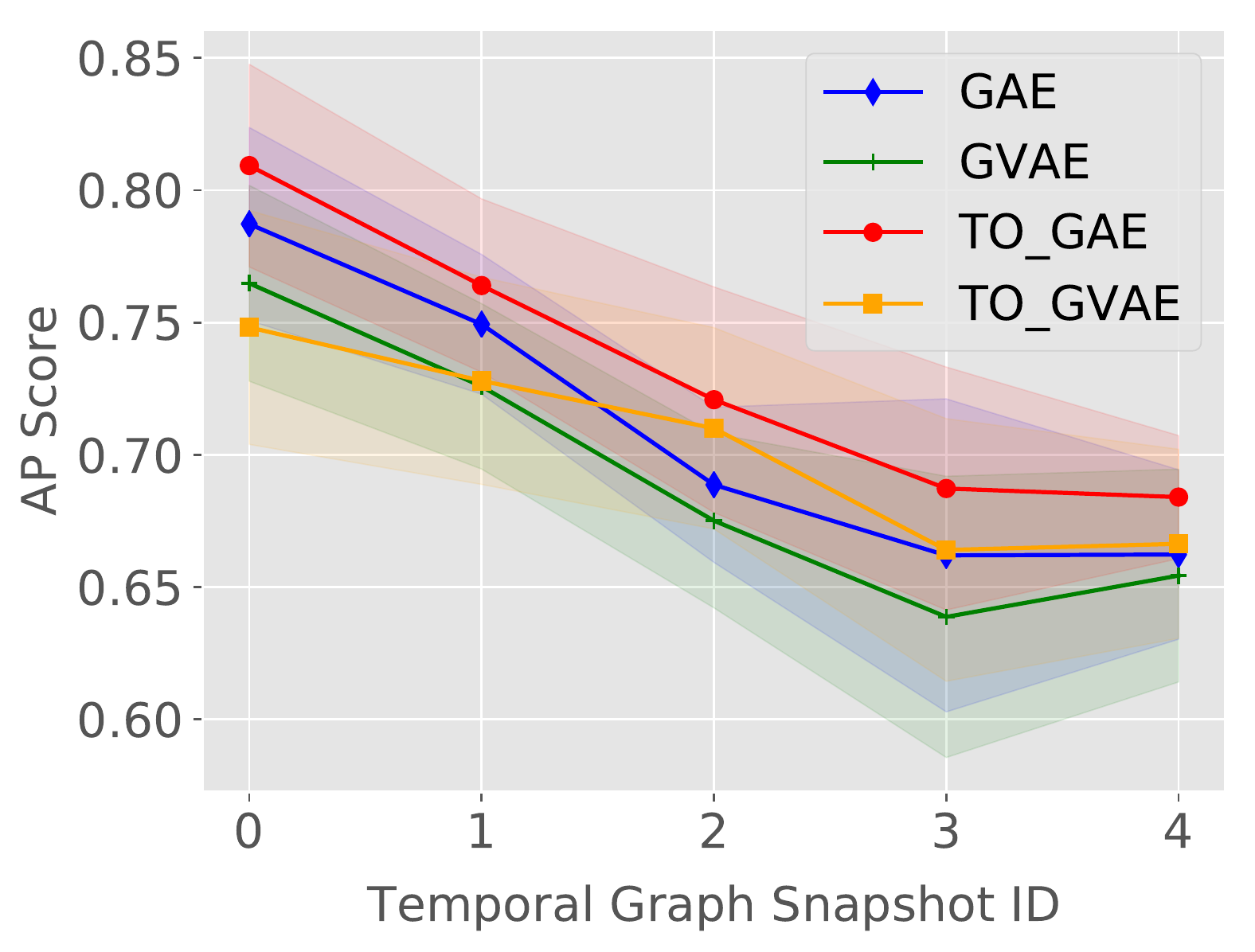}}
    \label{PRmiGN1}\hfill
  \subfloat[AP score on new edges]{%
      \includegraphics[width=0.25\linewidth]{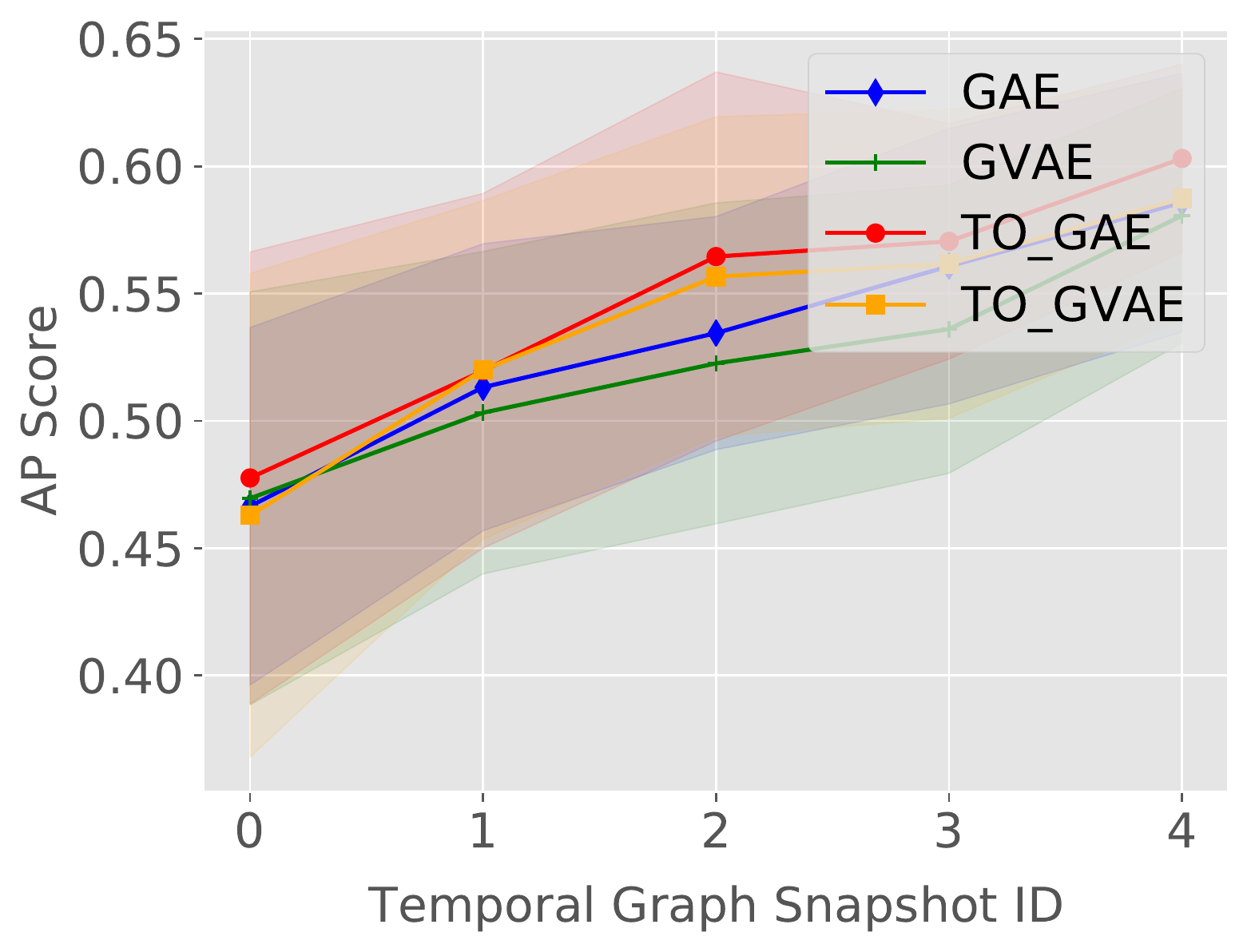}}
    \label{PRmiWI1}\\
  \caption{AUC and AP scores on the Cora dataset evolved via the configuration method with a 25\% chance of edges being rewired per time step. Values are presented for the whole graph and only on new edges which have been altered since the graph used for training. }
  \label{fig:config_25_cora} 
  \end{figure*}
  
  \begin{figure*} 
    \centering
  \subfloat[AUC score]{%
      \includegraphics[width=0.25\linewidth]{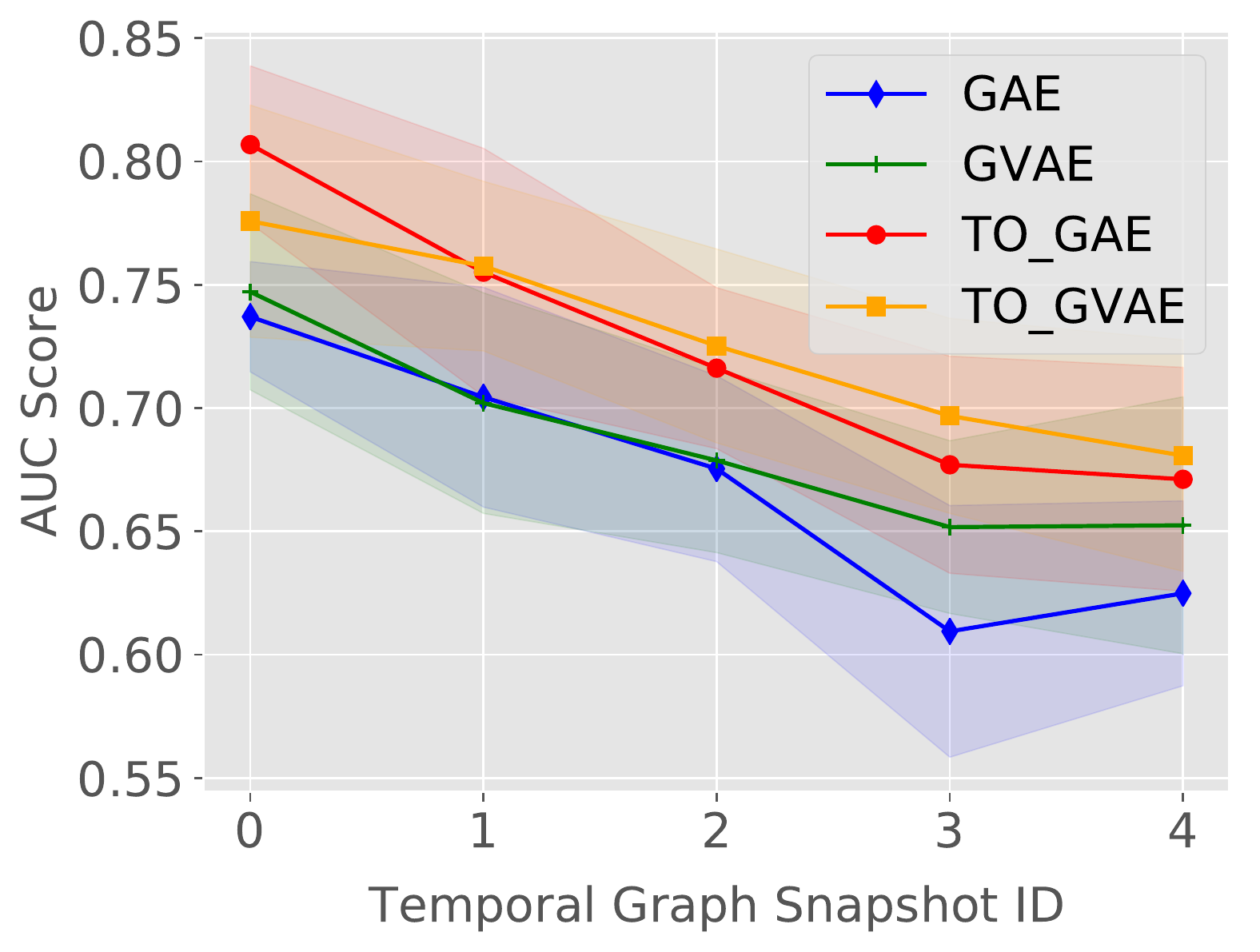}}
    \label{PRmiCA2}\hfill
  \subfloat[AUC score on new edges]{%
      \includegraphics[width=0.25\linewidth]{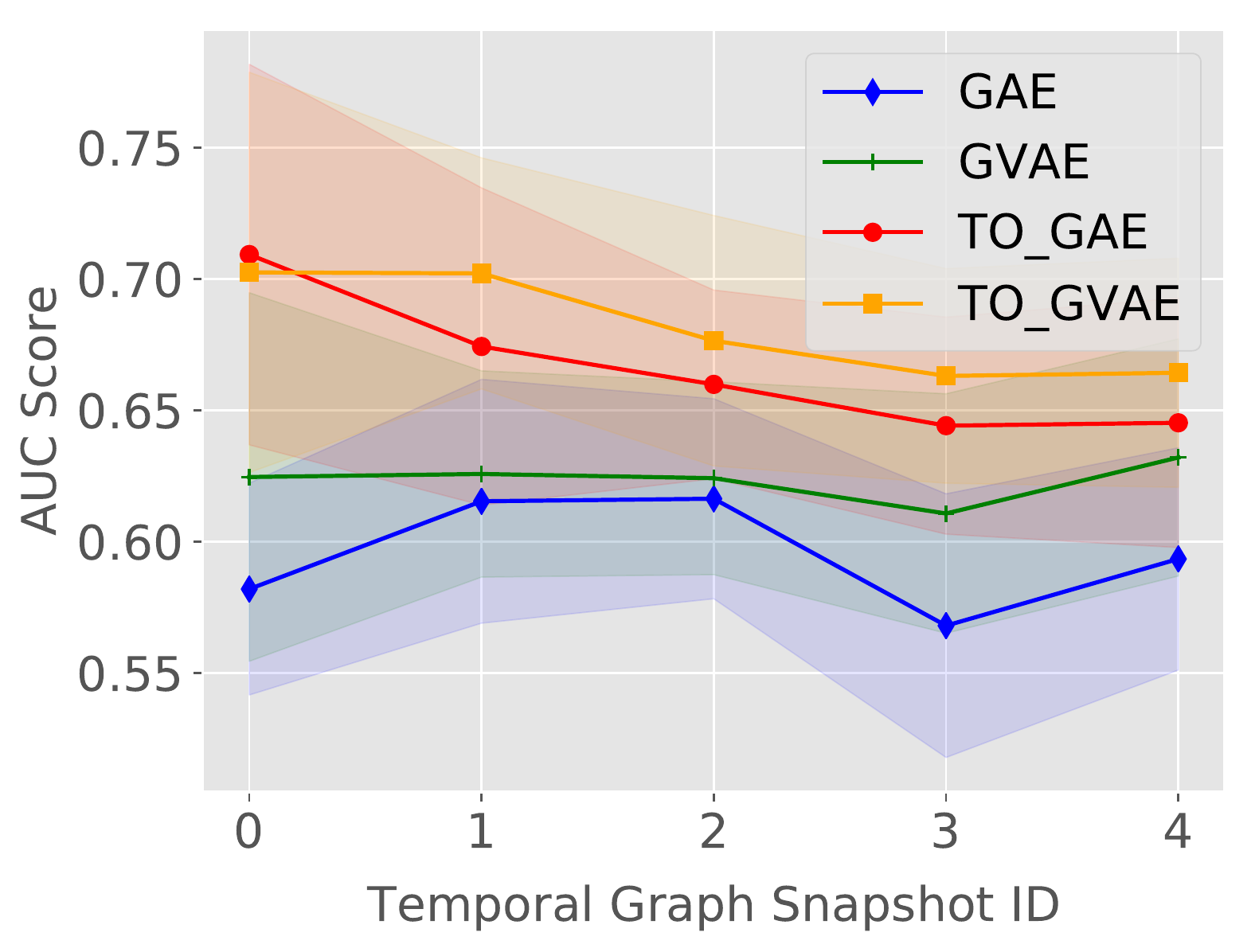}}
    \label{PRmiFB2}\hfill
  \subfloat[AP score]{%
      \includegraphics[width=0.25\linewidth]{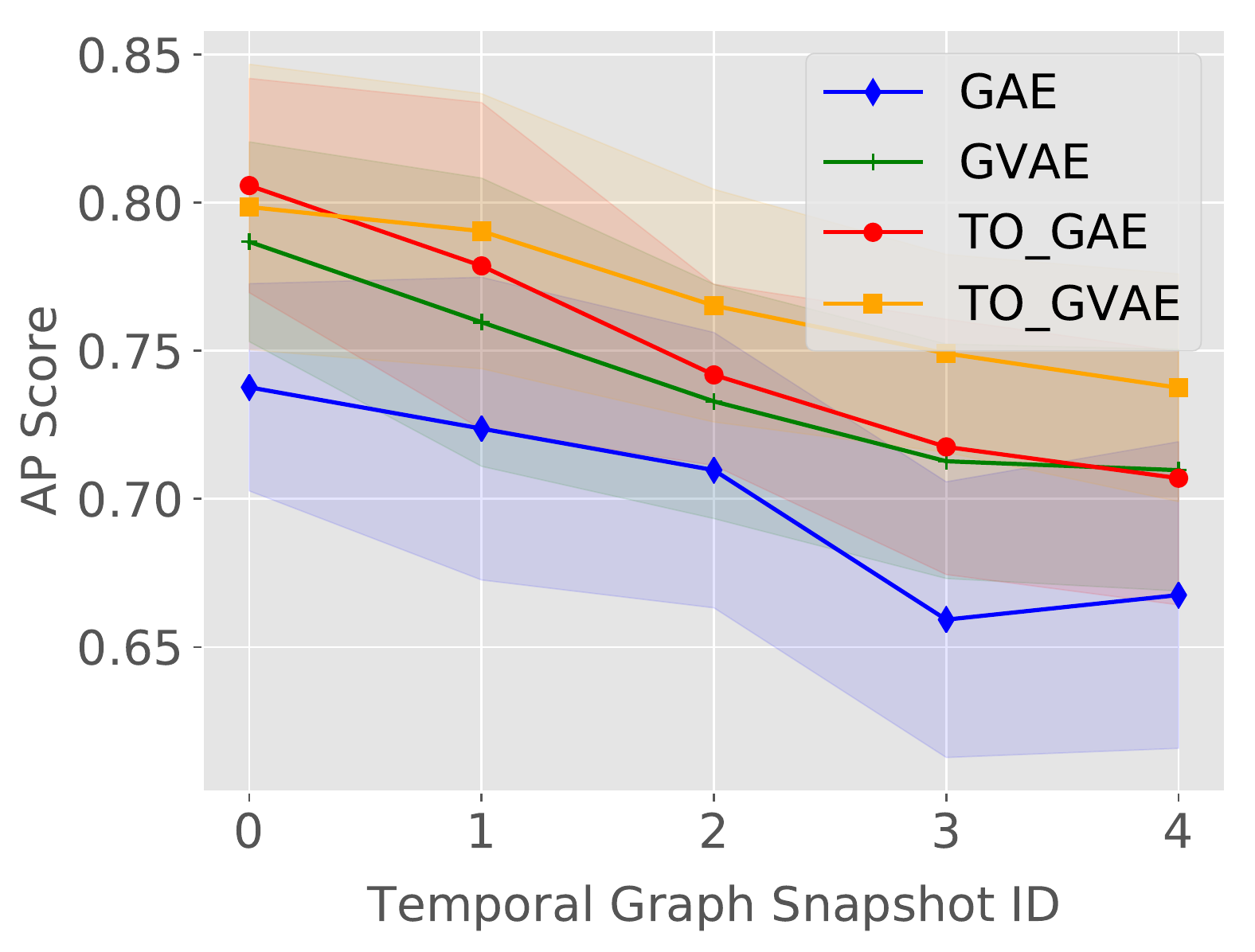}}
    \label{PRmiGN2}\hfill
  \subfloat[AP score on new edges]{%
      \includegraphics[width=0.25\linewidth]{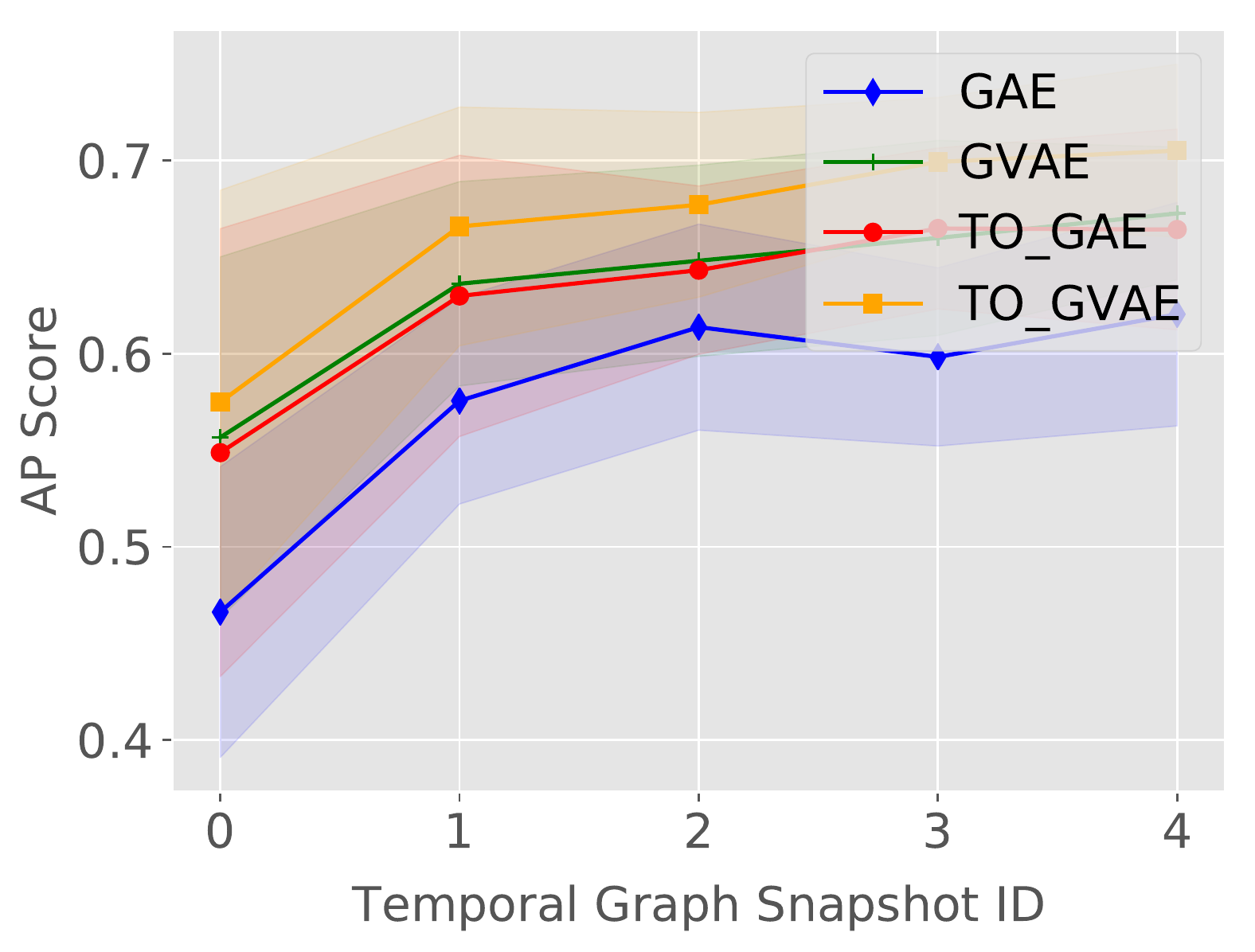}}
    \label{PRmiWI2}\\
  \caption{AUC and AP scores on the Citeseer dataset evolved via the configuration method with a 25\% chance of edges being rewired per time step. Values are presented for the whole graph and only on new edges which have been altered since the graph used for training. }
  \label{fig:config_25_citeseer} 
  \end{figure*}

\subsection{Datasets}
\label{sec:datasets}

We make use of the following empirical graph datasets presented in Table \ref{tab:datasets} when performing our experimental evaluation. For the two datasets which contain no inherent temporal information, we generate a synthetic evolutionary trajectory for the graph using one of the rewire processes detailed in section \ref{sec:randomrewire}. For the empirical cit-HepPh dataset, we create six snapshots based on a linear partitioning of the graph's timeline. 

\begin{table}[!t]
    \caption{Empirical Graph Datasets}
    \label{tab:datasets}
    \centering
    \begin{tabular}{c c c c c}
    \toprule
    \textbf{Dataset} & $|V|$ & $|E|$ & Temporal & Reference \T\B \\
    \midrule \midrule
    cit-HepPh & 34,546 & 421,578 & Snapshots & \cite{snapnets} \T \\
    
    cora & 2,708 & 5,429 & Synthetic & \cite{kipf2016variational}\;\\
    
    citeseer & 3,327 & 3,327 & Synthetic & \cite{kipf2016variational} \B \\
    
    \bottomrule
    \end{tabular}
  \end{table}

  \begin{figure*} 
    \centering
  \subfloat[AUC score]{%
      \includegraphics[width=0.25\linewidth]{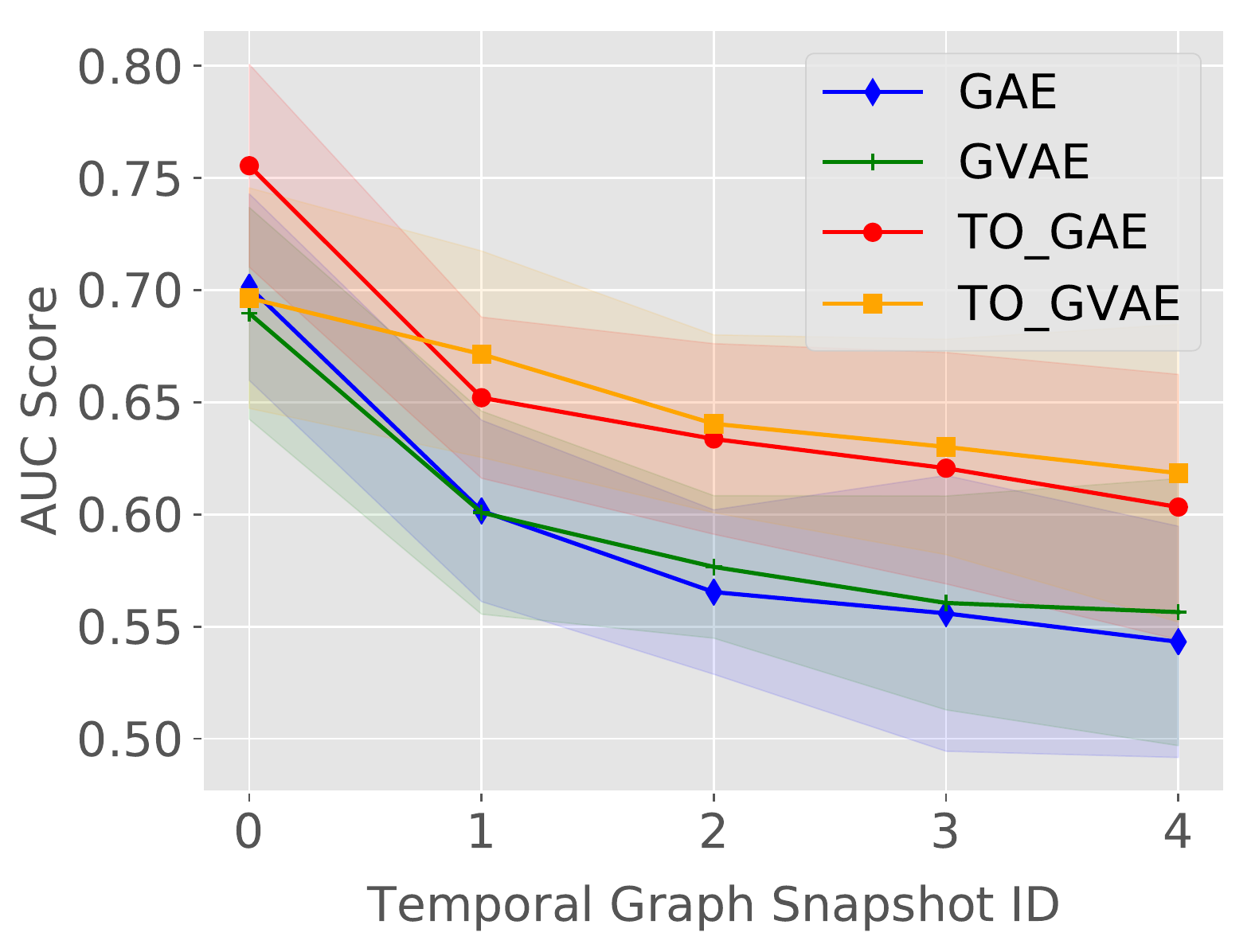}}
    \label{PRmiCA3}\hfill
  \subfloat[AUC score on new edges]{%
      \includegraphics[width=0.25\linewidth]{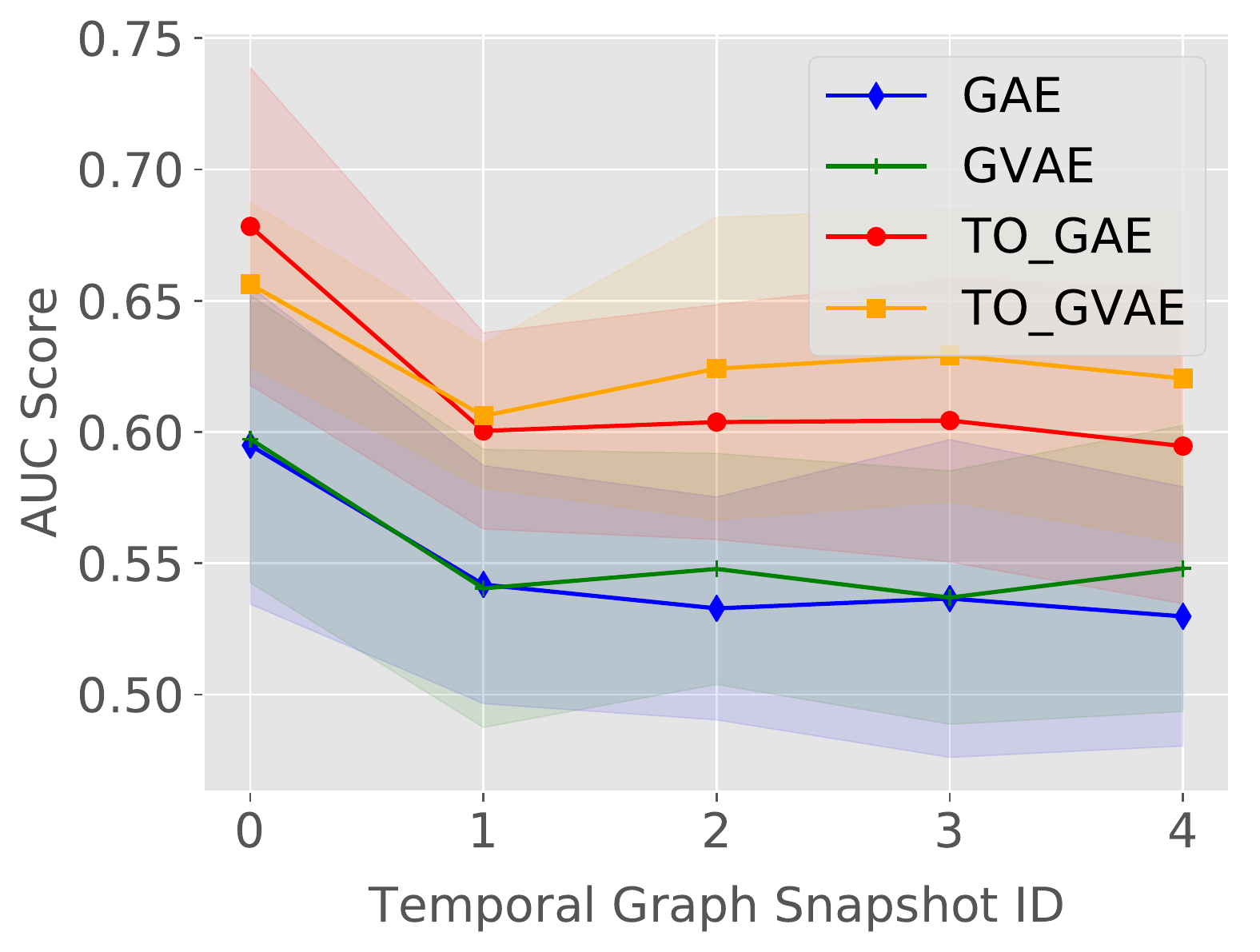}}
    \label{PRmiFB3}\hfill
  \subfloat[AP score]{%
      \includegraphics[width=0.25\linewidth]{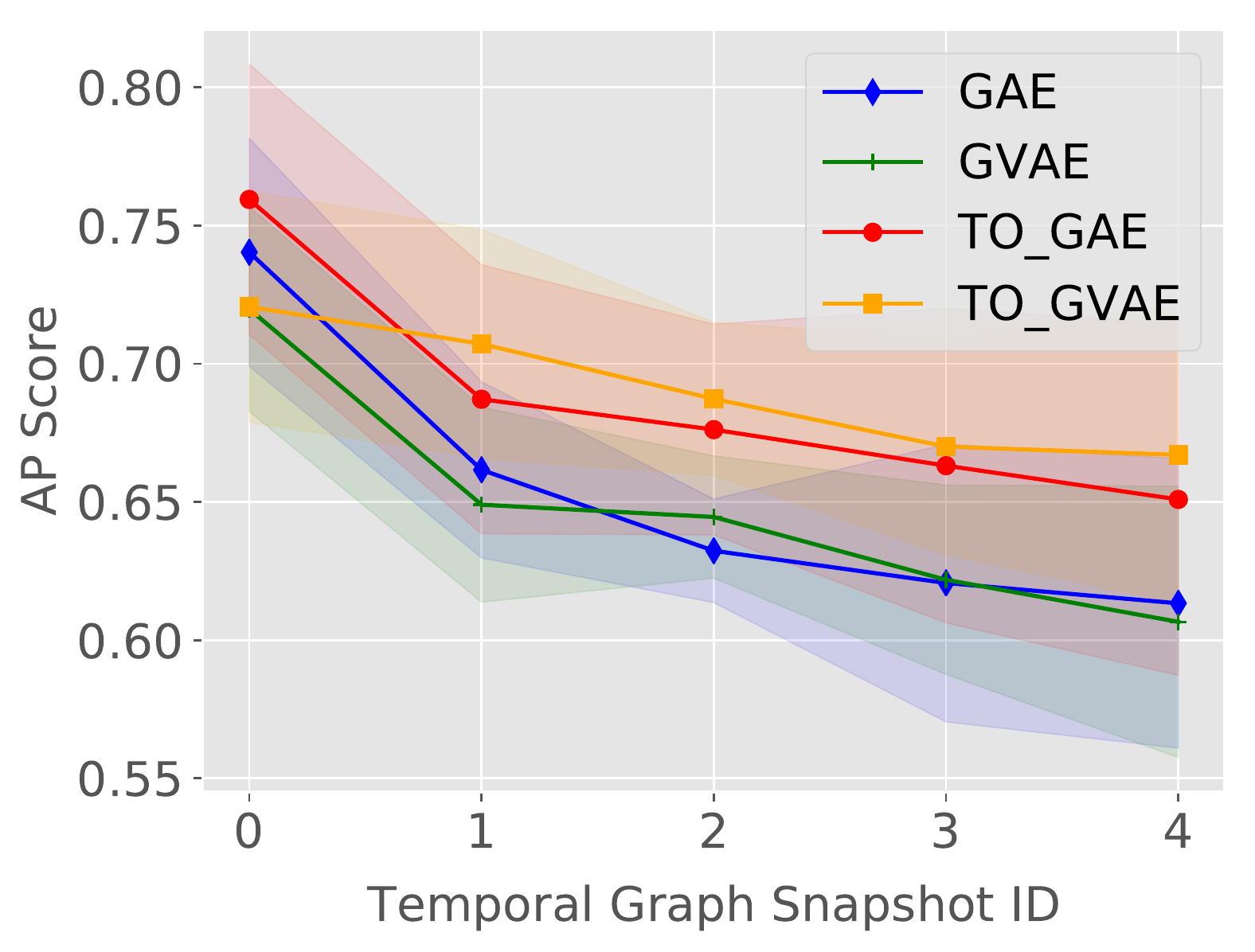}}
    \label{PRmiGN3}\hfill
  \subfloat[AP score on new edges]{%
      \includegraphics[width=0.25\linewidth]{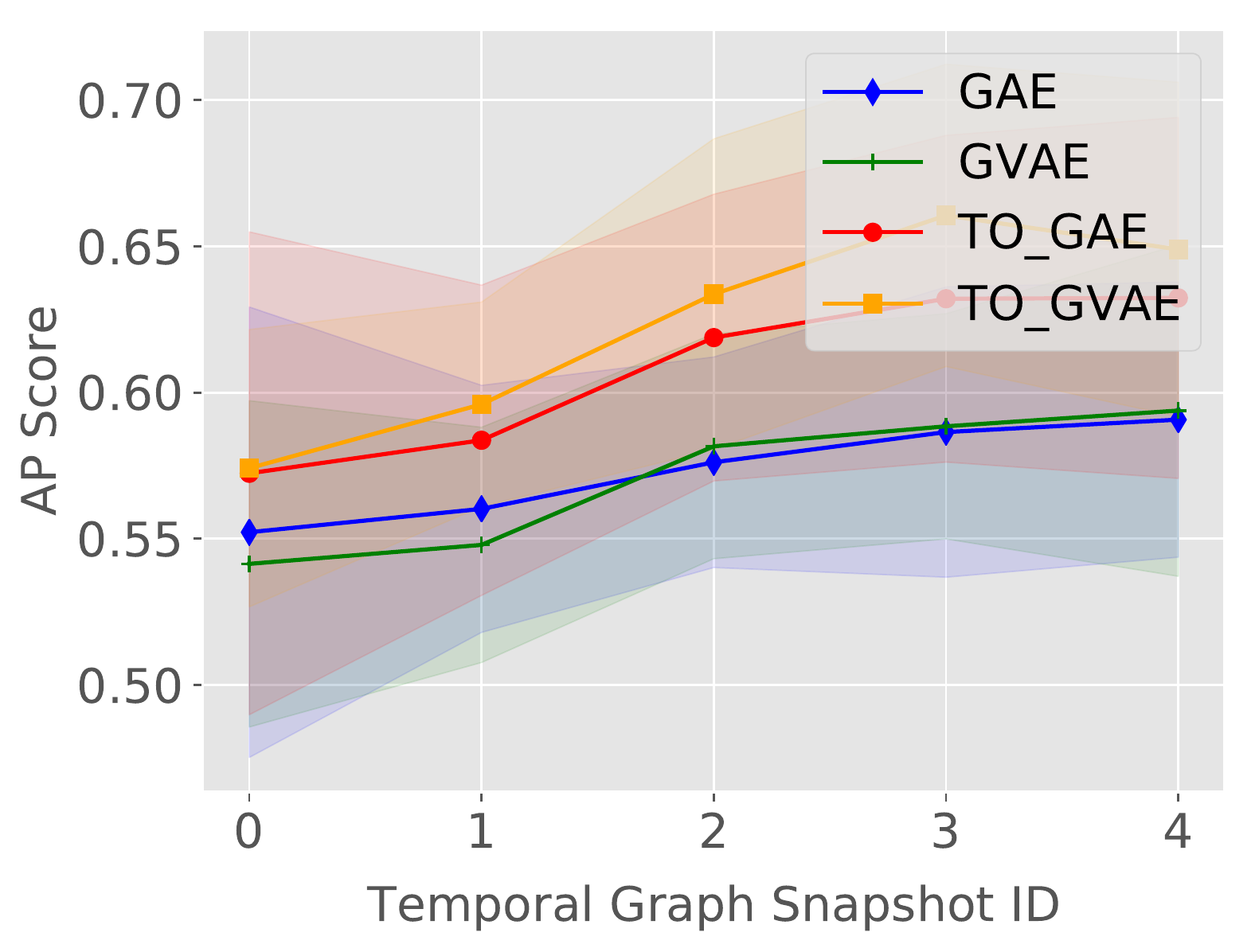}}
    \label{PRmiWI3}\\
  \caption{AUC and AP scores on the Cora dataset evolved via the configuration method with a 50\% chance of edges being rewired per time step. Values are presented for the whole graph and only on new edges which have been altered since the graph used for training. }
  \label{fig:config_50_cora} 
  \end{figure*}
  
  \begin{figure*} 
    \centering
  \subfloat[AUC score]{%
      \includegraphics[width=0.25\linewidth]{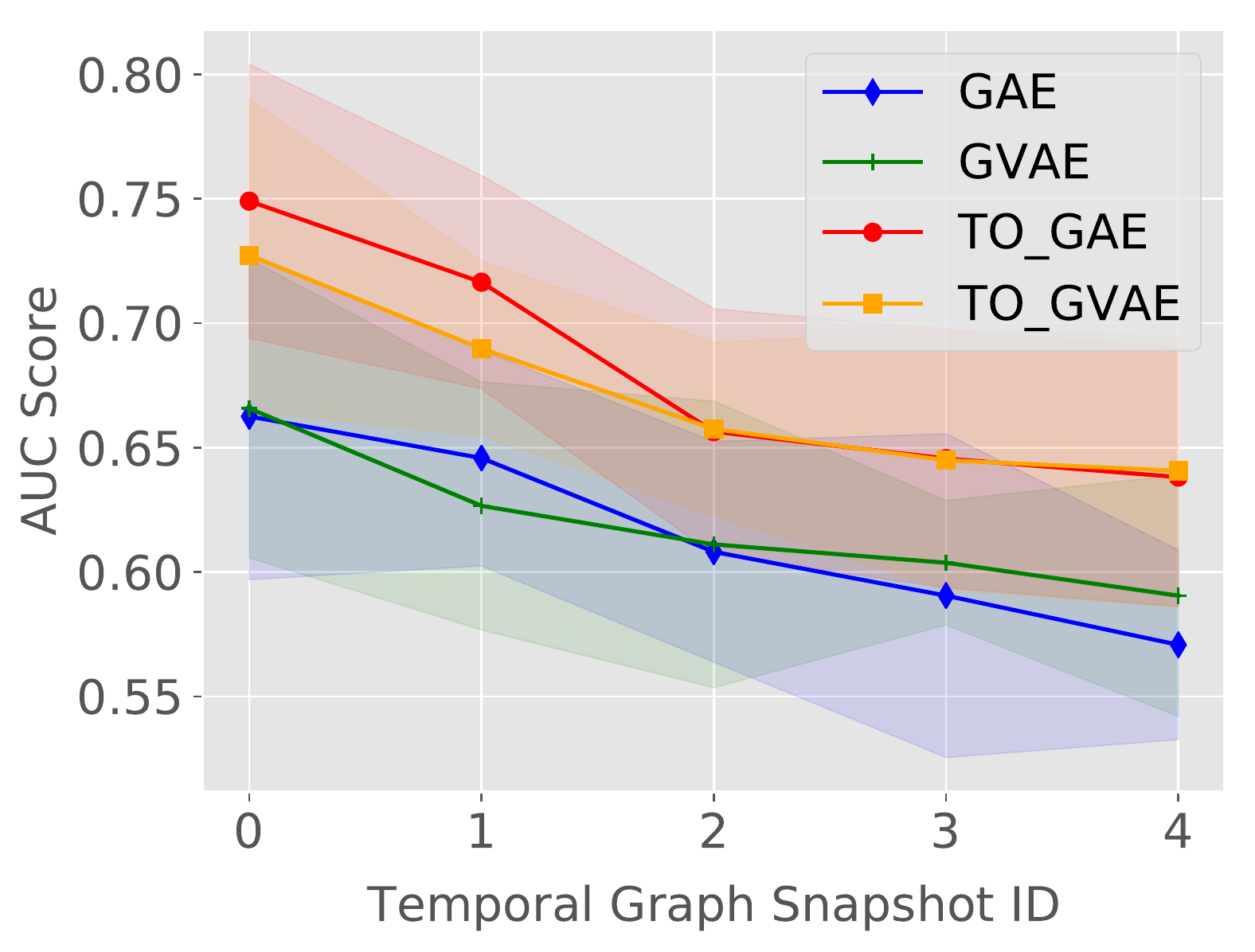}}
    \label{PRmiCA4}\hfill
  \subfloat[AUC score on new edges]{%
      \includegraphics[width=0.25\linewidth]{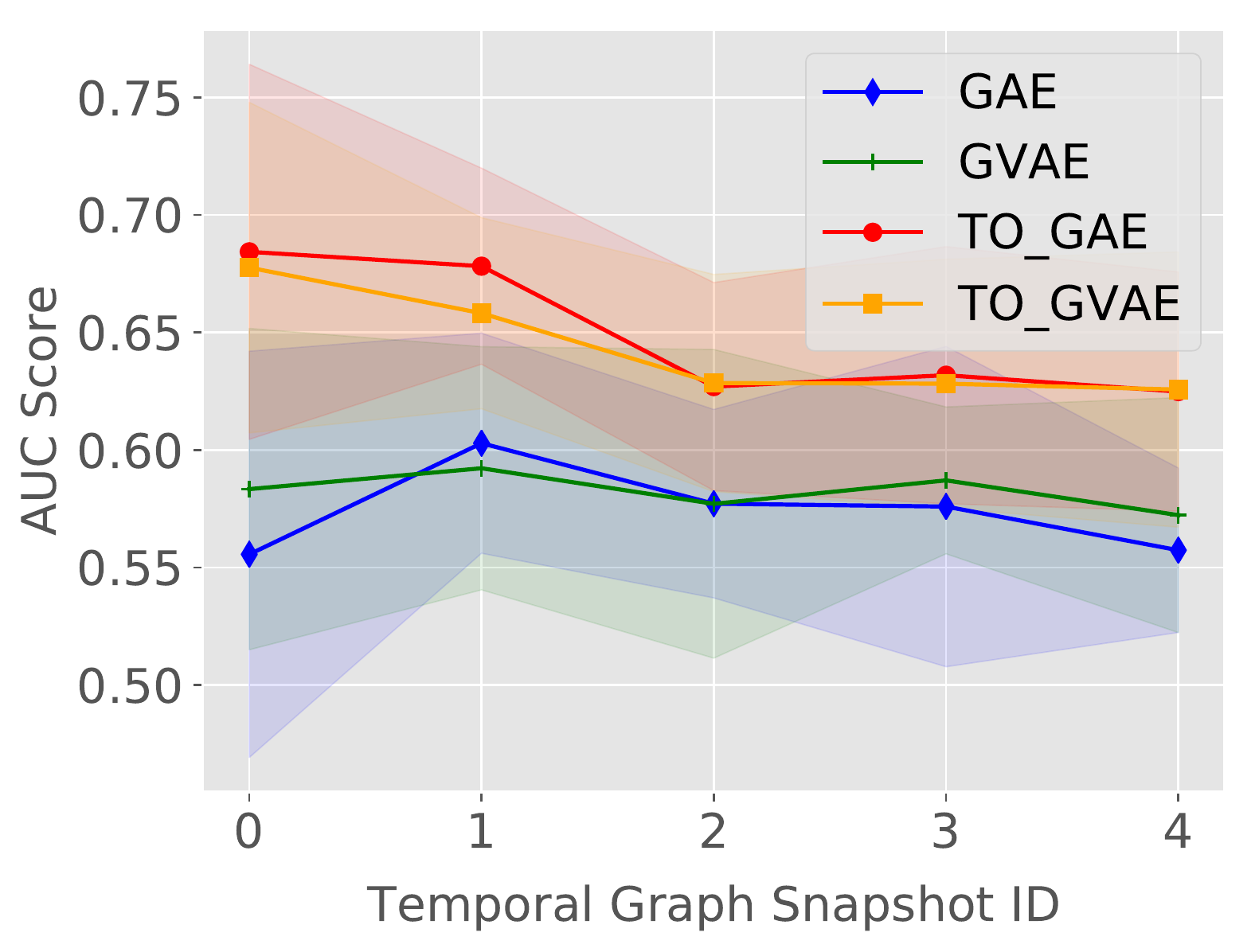}}
    \label{PRmiFB4}\hfill
  \subfloat[AP score]{%
      \includegraphics[width=0.25\linewidth]{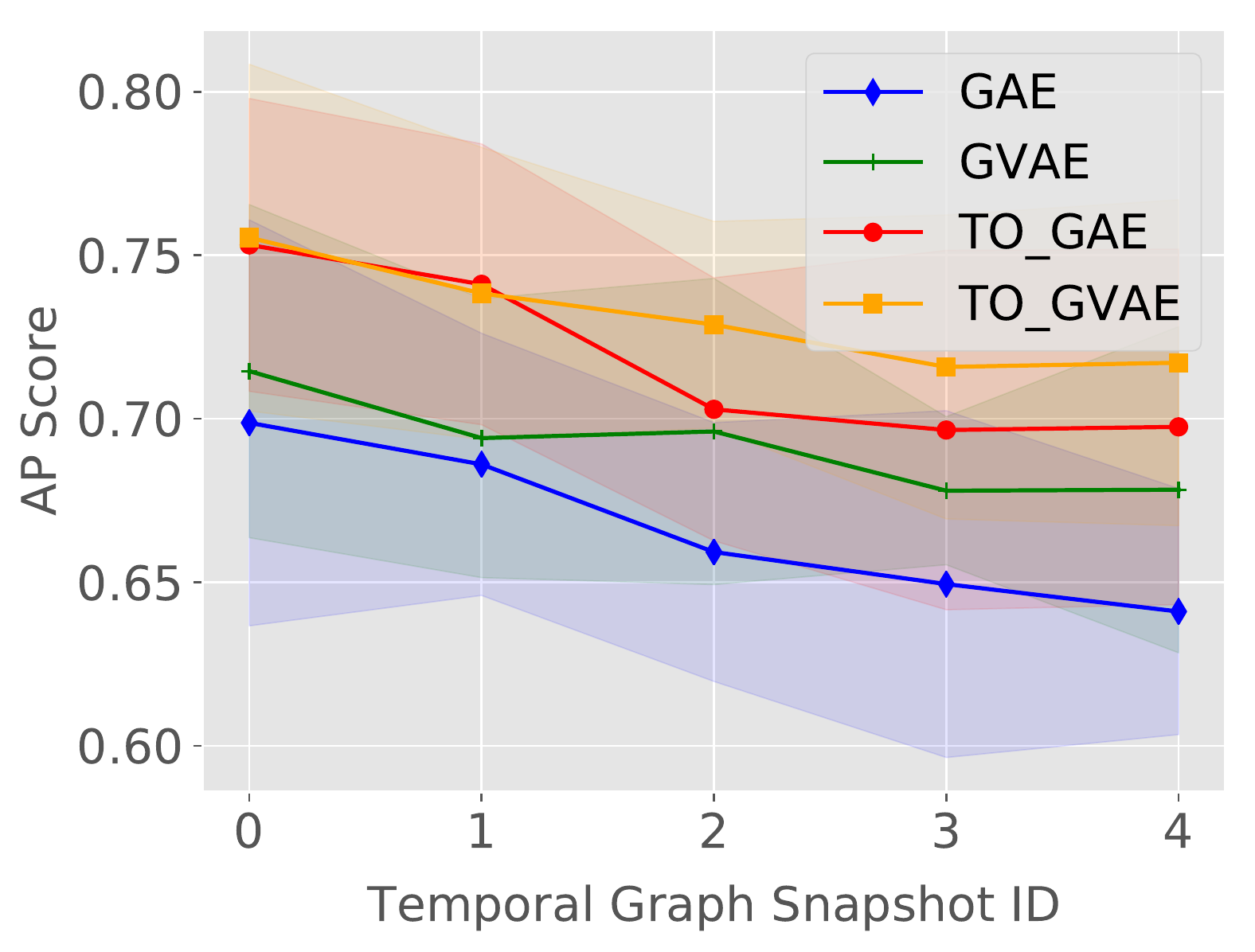}}
    \label{PRmiGN4}\hfill
  \subfloat[AP score on new edges]{%
      \includegraphics[width=0.25\linewidth]{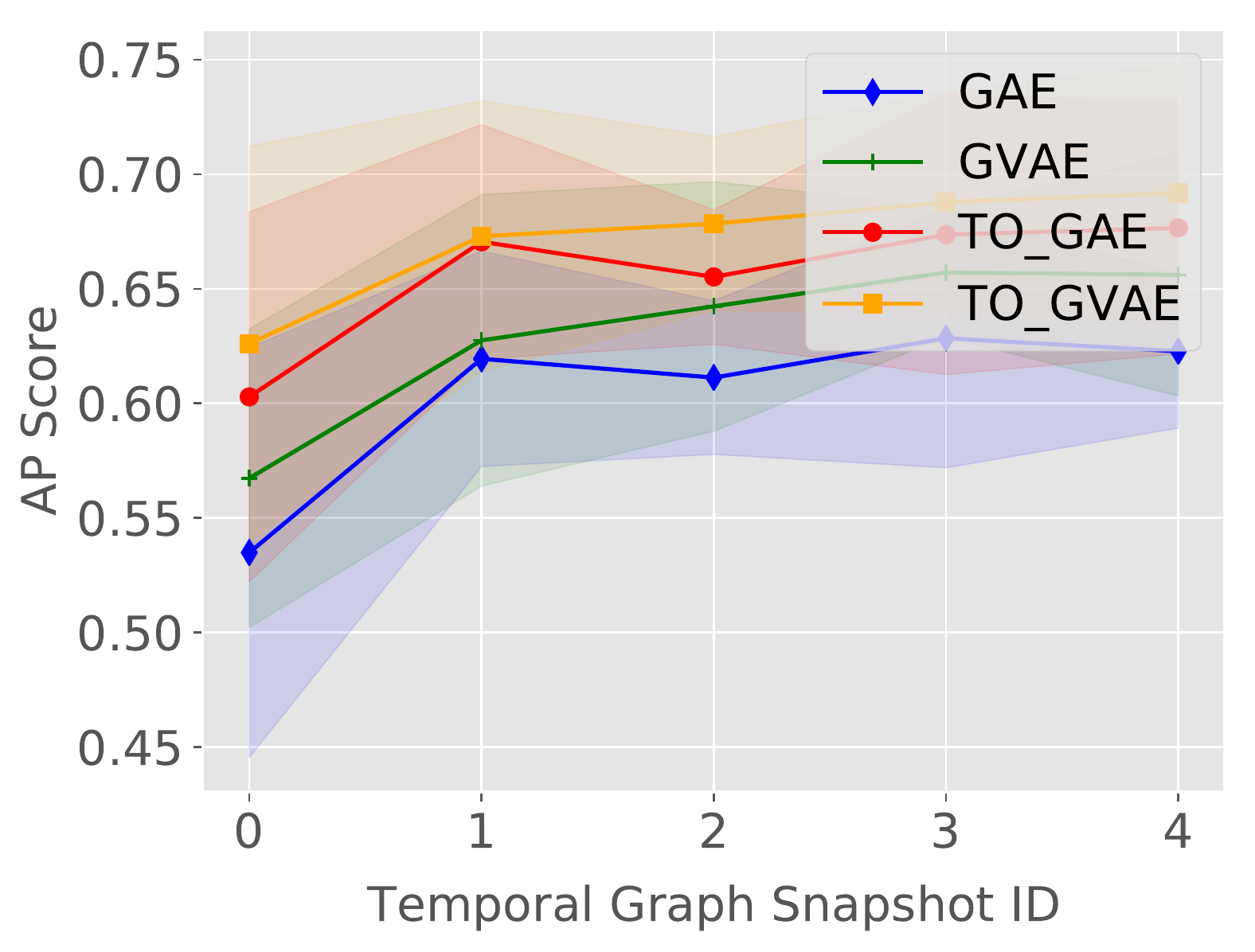}}
    \label{PRmiWI4}\\
  \caption{AUC and AP scores on the Citeseer dataset evolved via the configuration method with a 50\% chance of edges being rewired per time step. Values are presented for the whole graph and only on new edges which have been altered since the graph used for training. }
  \label{fig:config_50_citeseer} 
  \end{figure*}

\subsection{Random Rewire Process}
\label{sec:randomrewire}

In order to have large amounts of temporal graph data which has been evolved via a controllable process, we make use of the random rewire methodology \cite{bonner2016deep, bonner2016efficient}. The rewire process alters a given source graph's degree distribution by randomly altering the source and target of a set number of edges. During this re-wire process, it is not guaranteed that the source or target of the edge will be altered, indeed it is not always possible due to the graphs topology. Also, the rewiring process does not change the total number of edges or vertices within the graph. We employ two types of random rewire in this work:

\begin{itemize}

  \item \emph{Erd\H{o}s} - The edges are rewired such that the resulting topology of the graph begins to resemble a Erd\H{o}s-R\'{e}nyi graph, were edges are uniformly distributed between vertices.  
  
  \item \emph{Configuration} - The edges are rewired in such a way that each vertex approximately preserves it's associated number of edges, creating graphs with a similar degree distribution to the original.
  
\end{itemize}

\subsection{Baseline Approaches}
\label{sec:baselines}

Graph convolutional auto-encoders have been shown to have state-of-the-art performance in the link-prediction task for high-dimensional graphs \cite{kipf2016variational}. As such, we compare our approach to both non-probabilistic and variational versions of the convolutional graph auto-encoder, these will be denoted as GAE and GVAE in the results section. These approaches are trained as originally detailed by the authors \cite{kipf2016variational}, meaning that temporal information is not utilised. 

% \subsection{Experimental Environment}
% %TODO: Maybe delete
% Experimentation was performed on a compute system with 2 NVIDIA Tesla K40c's, 2.3GHz Intel Xeon E5-2650 v3, 64GB RAM and the following software stack: Ubuntu Server 18.04 LTS, Python 3.7, CUDA 9.0, CuDNN v7, PyTorch 0.4.1, scikit-learn 0.19.0, Graph-Tool 2.27 and NetworkX 2.0.

%----------------------------------------------------------------------------------------
%	SECTION - Results
%----------------------------------------------------------------------------------------
\section{Results}
\label{sec:results}

This section will present the results of the experimental evaluation. As outlined in Section \ref{sec:testingmeth}, we are testing two ways to evaluate the models: evolution pattern prediction and future link prediction. We present results on the datasets introduced in Section \ref{sec:datasets}, which contain both simulated and empirical evolving graph datasets. All the results presented are the mean, with the standard deviation, of ten repeats of the evaluation procedure, using a random train/test split for each repetition. 

\subsection{Simulated Graph Evolution}

We first present results using time series simulated via the random rewire processes introduced in Section \ref{sec:randomrewire} on both the cora and citeseer datasets. For this experiment, we train using the original unaltered graph to reconstruct the next graph in the time series. We then measure the performance of the resulting embeddings at predicting edges in future graph snapshots using the two inference approaches outlined in Section \ref{sec:testingmeth}. 

\subsubsection{Evolution Pattern Prediction}

\begin{figure*}
  \centering
\subfloat[AUC score on Cora]{%
    \includegraphics[width=0.25\linewidth]{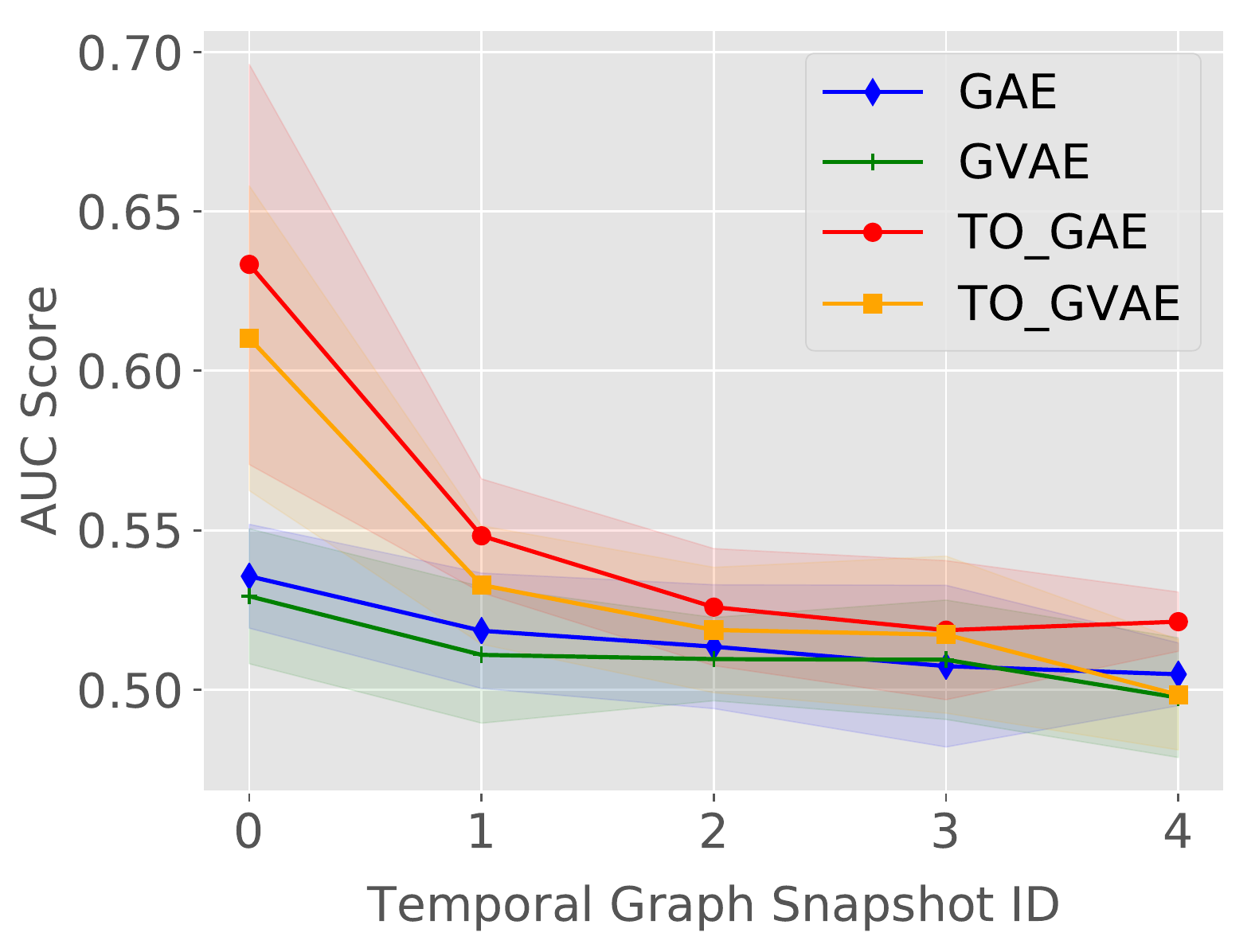}} 
  \label{PRmiCA5}\hfill
\subfloat[AP score on Cora]{%
    \includegraphics[width=0.25\linewidth]{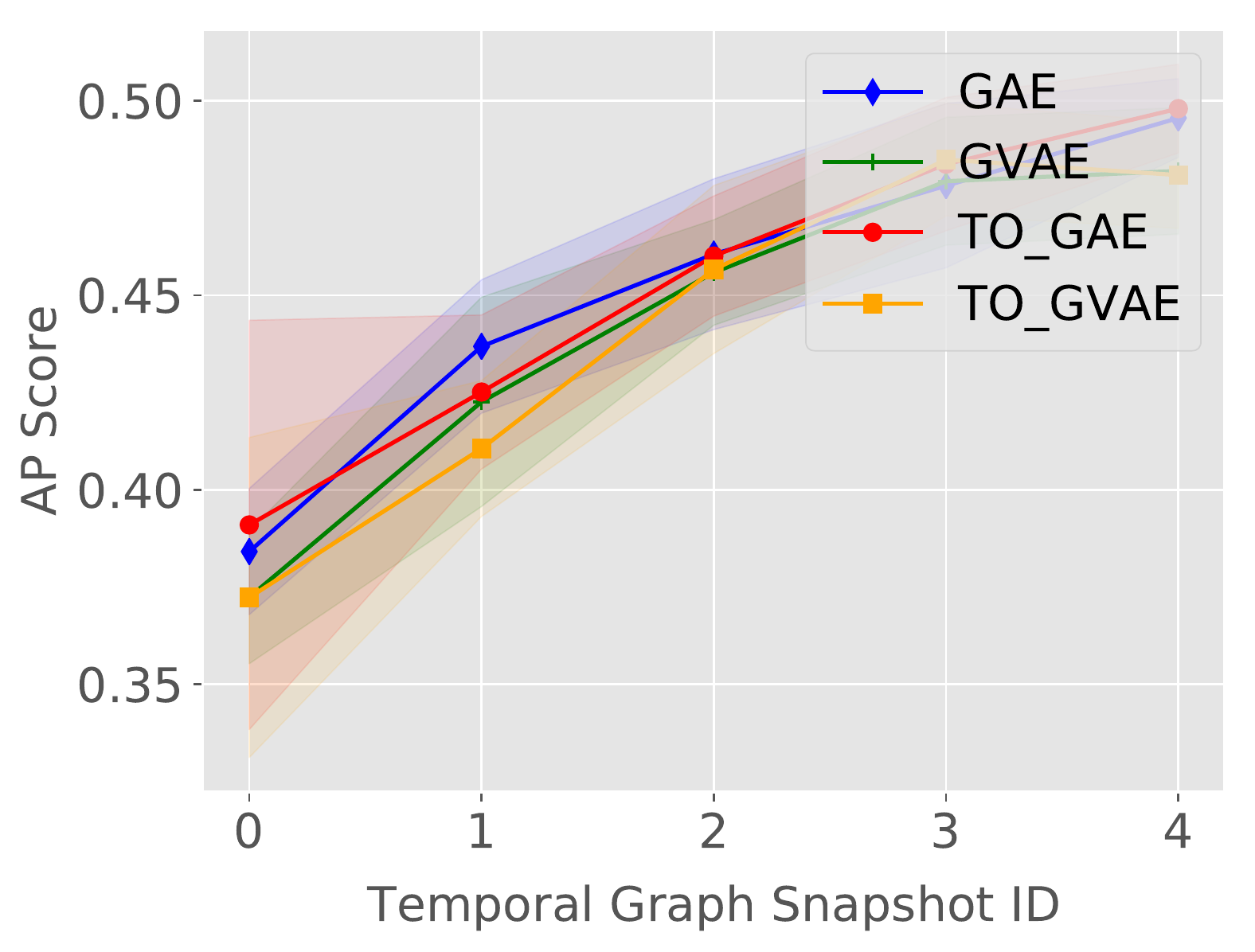}}
  \label{PRmiFB5}\hfill
\subfloat[AUC score on Citeseer]{%
    \includegraphics[width=0.25\linewidth]{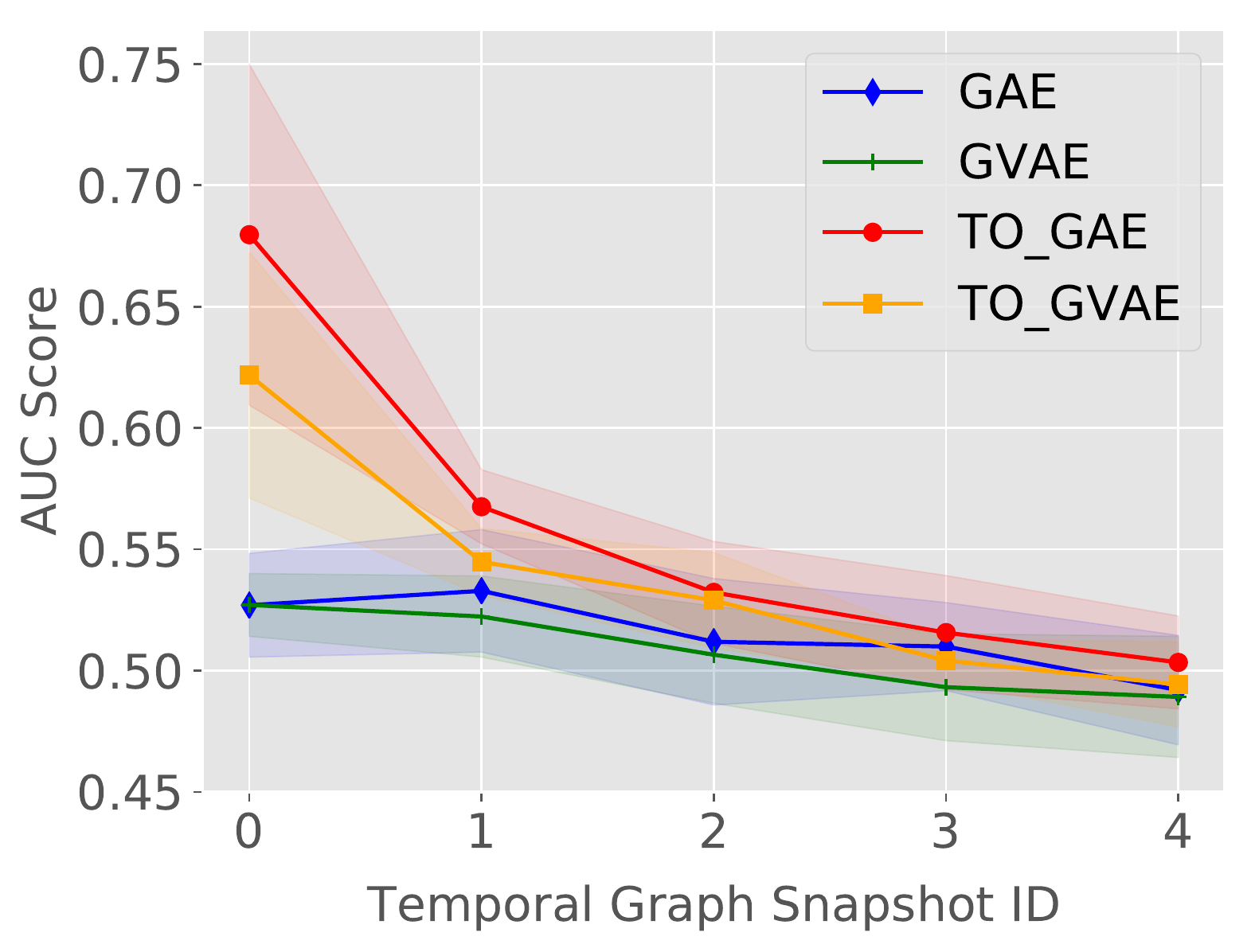}}
  \label{PRmiGN5}\hfill
\subfloat[AP score on Citeseer]{%
    \includegraphics[width=0.25\linewidth]{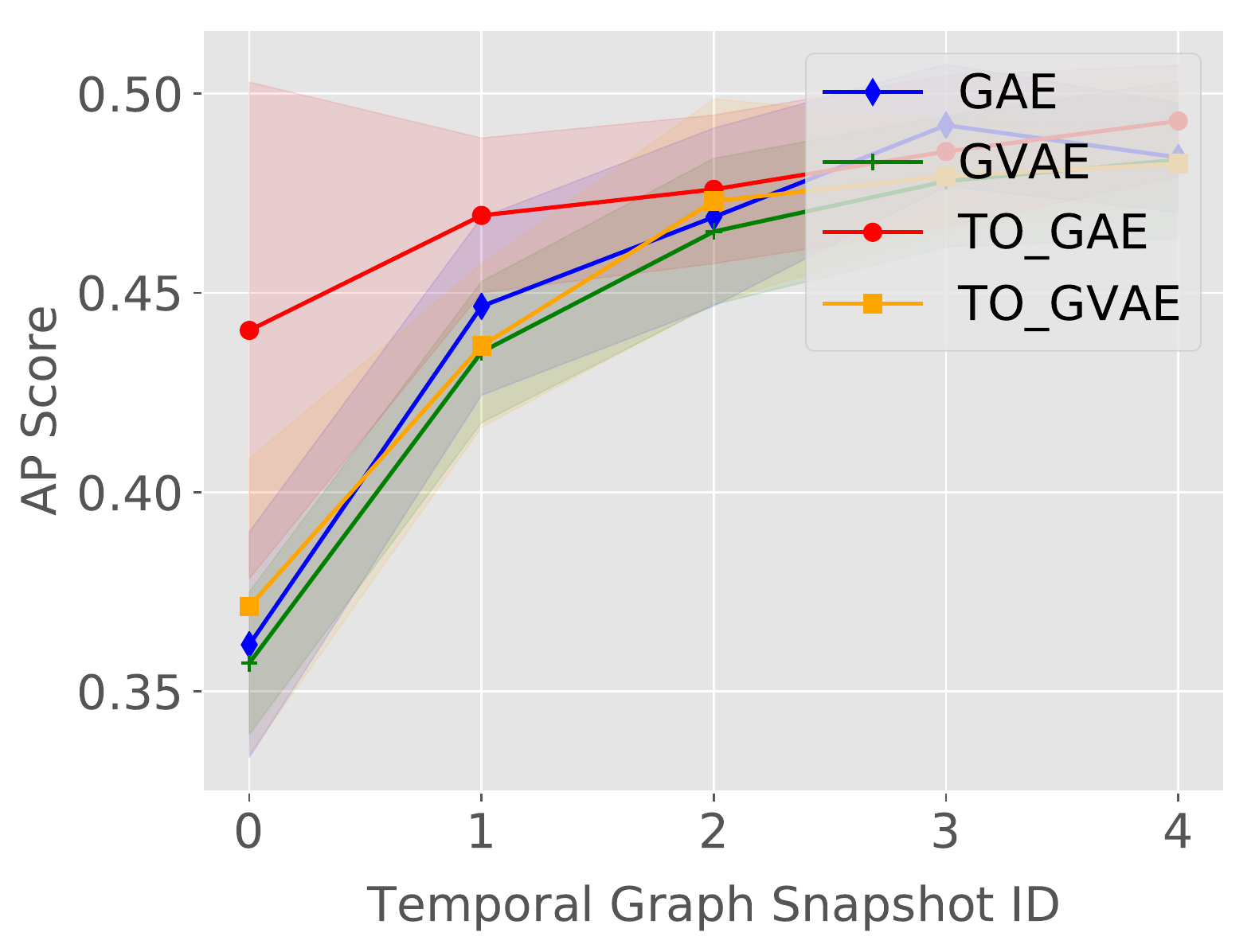}}
  \label{PRmiWI5}\\
\caption{AUC and AP sscoes for the future link prediction task on both the Cora and Citeseer datasets evolved using the Erd\H{o}s rewire method with $|E|/2$ edges having the chance of being rewired. The results presented are scores for predicting only new edges which have appeared after the original graph used for training the model.}
\label{fig:edros_change_50} 
\vskip -10pt
\end{figure*}

\begin{figure*}
  \centering
\subfloat[AUC score on Cora]{%
    \includegraphics[width=0.25\linewidth]{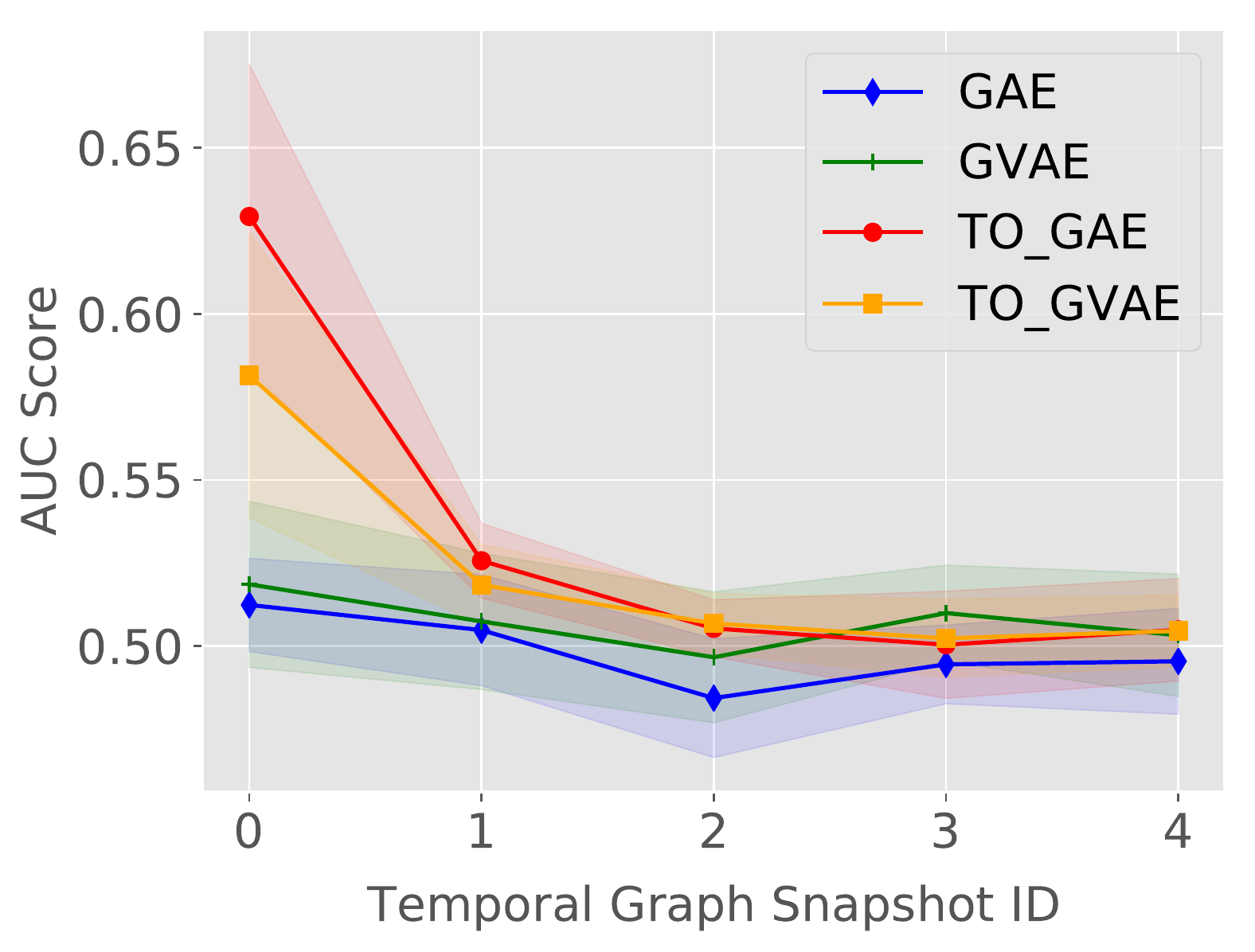}} 
  \label{PRmiCA6}\hfill
\subfloat[AP score on Cora]{%
    \includegraphics[width=0.25\linewidth]{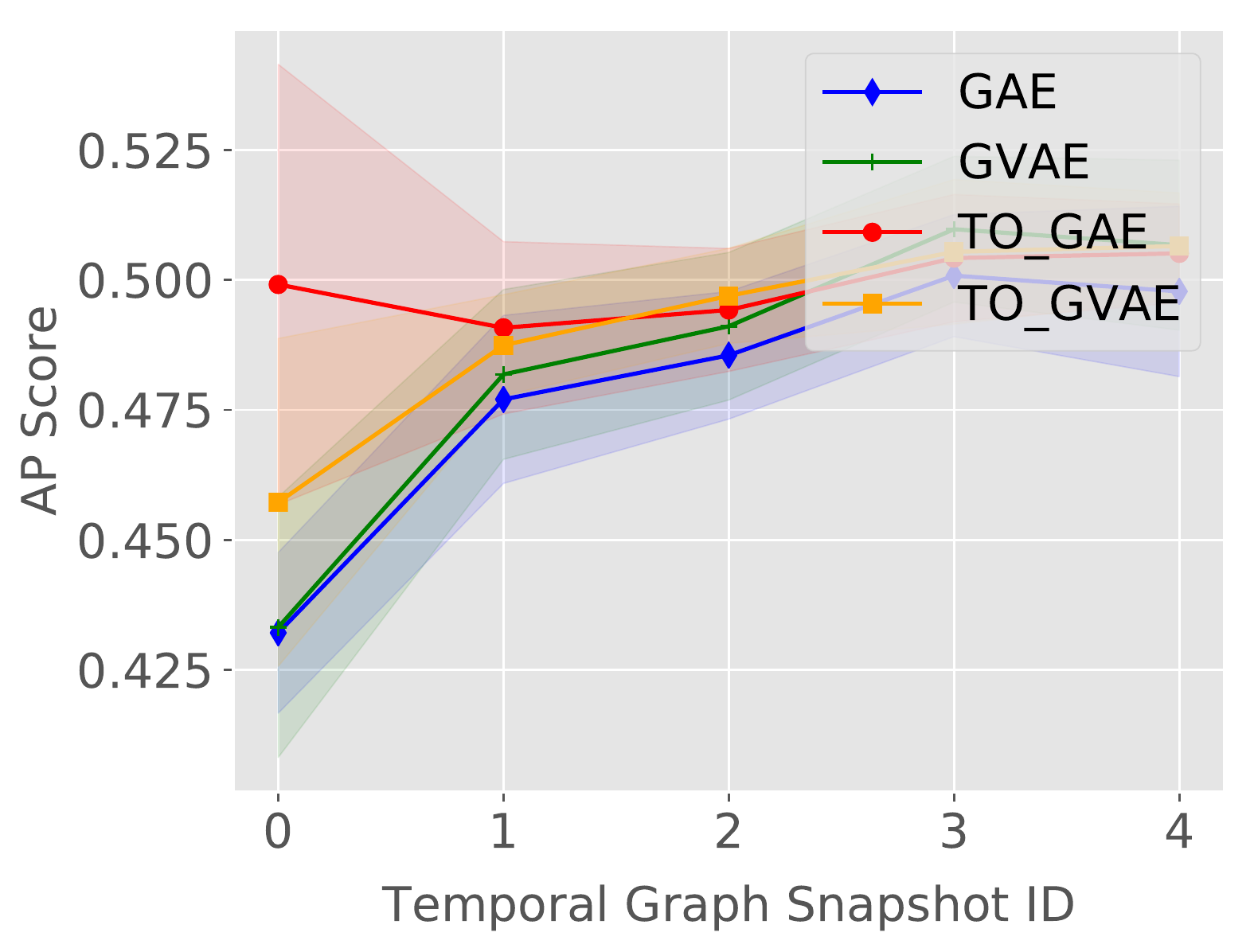}}
  \label{PRmiFB6}\hfill
\subfloat[AUC score on Citeseer]{%
    \includegraphics[width=0.25\linewidth]{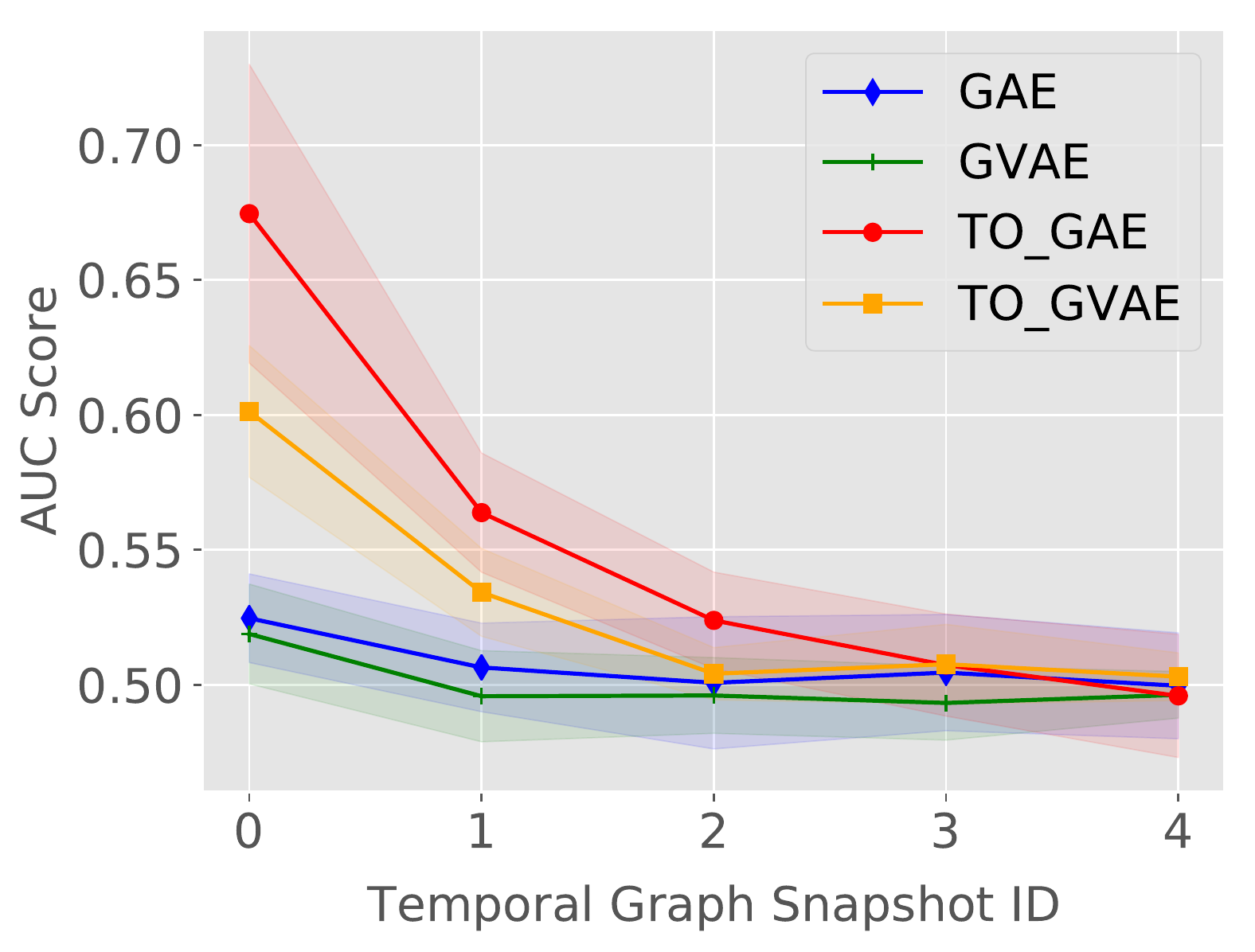}}
  \label{PRmiGN6}\hfill
\subfloat[AP score on Citeseer]{%
    \includegraphics[width=0.25\linewidth]{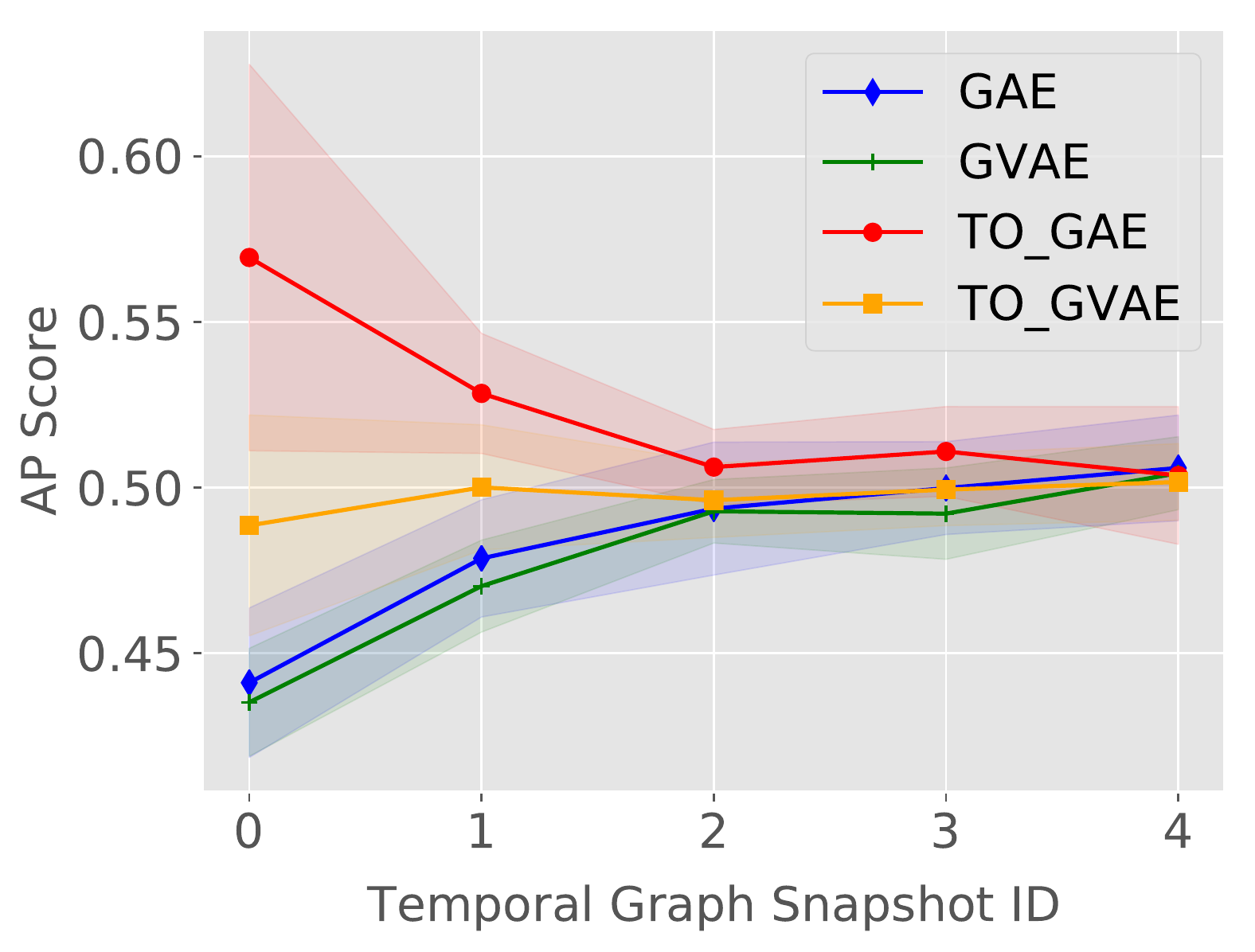}}
  \label{PRmiWI6}\\
\caption{AUC and AP scores for the future link prediction task on both the Cora and Citeseer datasets evolved using the Erd\H{o}s rewire method with the complete set of $E$ having the chance of being rewired. The results presented are scores for predicting only new edges which have appeared after the original graph used for training the model.} 
\label{fig:edros_change_100} 
\vskip -10pt
\end{figure*}

Figure \ref{fig:config_25_cora} shows how well the models perform upon the Cora dataset with smaller possible perturbations introduced by the rewire process. The figure shows how both the temporally offset methods we introduce generally demonstrate greater performance than the baseline approaches. When considering the AUC score on new edges, we can see a large increase in performance. Our methods also show performance above the baselines when reconstructing the full graph, demonstrating that the temporal offset process does not harm the abilty to predict edges which have not changed. 

  % ------------------------------------------- 25% -------------------------------------------

Figure \ref{fig:config_25_citeseer} contains the results for the Citeseer dataset, again using the smaller rewire probability. This figure continues the trends established in the previous figure, with the temporally offset methods beating the baseline methods. However, this time it is the variational approach which often demonstrates the greater performance, especially when only new edges are considered. 
  % ------------------------------------------- 50% -------------------------------------------

The next two sets of figures show results for when a large possible perturbation is made in-between each graph time step. Figure \ref{fig:config_50_cora} demonstrates that, as was expected, when even large steps are made between graphs, the gap between our approaches and the baselines also increases. The temporally offset methods show a clear increase in performance, even when using the model to make predictions about graphs later in the time series, which will have a quite different topological structure to the graph used to train the model. 

Figure \ref{fig:config_50_citeseer} shows the results for the Citeseer dataset using the higher level of possible perturbation. The figure continues the trend of the previous results by showing the temporally offset models to be better at predicting future changes in the graph, even when considering both the unchanged and new edges.

\subsubsection{Future Link Prediction}

For assessing the ability of the various models to continue to make accurate link predictions as the graph undergoes heavy topological change, we make use of the Erd\H{o}s-based random rewire method. Figure \ref{fig:edros_change_50} shows the results for only new edges on both the cora and citeseer datasets when 50\% of the edges have the chance of being rewired in-between each graph snapshot. The figure shows that our temporally offset training method is more robust to the Erd\H{o}s rewired edges, as it displays a higher level of predictive performance particularly when regarding the AUC metric. Interesting, of the two temporally offset models, it is the non-probabilistic approach which displays greater performance across both datasets.

Figure \ref{fig:edros_change_100} highlights the performance on new edges when all the edges have the chance to be rewired between graph snapshots. The results show that the temporal offset approaches are generally more robust to the large change in graph topology between graph snapshots. With both demonstrating greater performance, especially at earlier points in the time series. Continuing the trend established in the previous experiment, the non-probabilistic temporally offset model out performs the variational approach. We hypothesise that as the variational approach is a more complex model, it is over fitting more strongly to the non-rewired original graph edges, making it less able to learn the Erd\H{o}s pattern of rewiring. 

\subsection{Empirical Time-Series}

\subsubsection{Evolution Pattern Prediction}

Table \ref{tab:HepPh} displays the results for the task of evolution pattern prediction for all models for the cit-HepPh dataset. The table shows that both the temporally offset methods significantly outperform the baseline approaches on both whole graph and new edges metrics at this particular task. The gap in performance between the temporally offset and normal approaches for this empirical dataset is larger than on the previous synthetic results, indicating that our approach is much better able to learn the temporal dynamics of real datasets. We can also see that the variational temporally offset approach is often the best performing of the two approaches, particularly at later time-points. 

\begin{table*}[h!]
  \centering
  %\resizebox{\textwidth}{!}{
  \begin{tabular}{ll c c c c c}
  \toprule
  \textbf{Model}  & \textbf{Metric} & $G_1$           & $G_2$         & $G_3$          & $G_4$         & $G_5$ \T\B \\
  \midrule \midrule

\multirow{4}{*}{GVAE}   & AUC  & $0.699(\pm0.0133)$  & $0.6327(\pm0.0036)$  & $0.5913(\pm0.0022)$  & $0.5821(\pm0.0049)$  & $0.5771(\pm0.0122)$  \\
                        & AP  & $0.8023(\pm0.0089)$  & $0.7459(\pm0.0056)$  & $0.7036(\pm0.001)$  & $0.6853(\pm0.0035)$  & $0.685(\pm0.0095)$  \\
                        & NE-AUC  & $0.5358(\pm0.0103)$  & $0.513(\pm0.0036)$  & $0.5012(\pm0.0098)$  & $0.5034(\pm0.0067)$  & $0.4979(\pm0.0112)$  \\
                        & NE-AP  & $0.5875(\pm0.0071)$  & $0.5698(\pm0.0055)$  & $0.5663(\pm0.0067)$  & $0.5598(\pm0.0079)$  & $0.552(\pm0.0111)$  \\

\midrule
\multirow{4}{*}{GAE}  & AUC  & $0.653(\pm0.0159)$  & $0.5939(\pm0.0062)$  & $0.5546(\pm0.0034)$  & $0.5343(\pm0.0043)$  & $0.5343(\pm0.0043)$  \\
                      & AP  & $0.7817(\pm0.0103)$  & $0.7293(\pm0.0069)$  & $0.6875(\pm0.0023)$  & $0.6643(\pm0.006)$  & $0.6643(\pm0.006)$  \\
                      & NE-AUC  & $0.4665(\pm0.0172)$  & $0.4512(\pm0.0033)$  & $0.4526(\pm0.0041)$  & $0.4436(\pm0.0031)$  & $0.4436(\pm0.0031)$  \\
                      & NE-AP  & $0.5541(\pm0.0131)$  & $0.5416(\pm0.0055)$  & $0.5434(\pm0.0031)$  & $0.5286(\pm0.0077)$  & $0.5286(\pm0.0077)$  \\

\midrule
\midrule
\multirow{4}{*}{TO-GVAE}  & AUC  & $0.9943(\pm0.0004)$  & $\mathbf{0.873(\pm0.0022)}$  & $\mathbf{0.7728(\pm0.0031)}$  & $\mathbf{0.726(\pm0.0084)}$  & $\mathbf{0.7281(\pm0.0045)}$  \\
                      & AP  & $\mathbf{0.9925(\pm0.0011)}$  & $\mathbf{0.9197(\pm0.0015)}$  & $\mathbf{0.8515(\pm0.0027)}$  & $\mathbf{0.8158(\pm0.0056)}$  & $\mathbf{0.8177(\pm0.0034)}$  \\
                      & NE-AUC  & $0.995(\pm0.001)$  & $\mathbf{0.8203(\pm0.0042)}$  & $\mathbf{0.7076(\pm0.0016)}$  & $\mathbf{0.6641(\pm0.0089)}$  & $\mathbf{0.6591(\pm0.008)}$  \\
                      & NE-AP  & $\mathbf{0.989(\pm0.003)}$  & $\mathbf{0.8615(\pm0.0043)}$  & $\mathbf{0.776(\pm0.0023)}$  & $\mathbf{0.7396(\pm0.0055)}$  & $\mathbf{0.7367(\pm0.0044)}$  \\

\midrule
\multirow{4}{*}{TO-GAE}   & AUC  & $\mathbf{0.9944(\pm0.0012)}$  & $0.8702(\pm0.0029)$  & $0.7629(\pm0.0062)$  & $0.711(\pm0.0095)$  & $0.711(\pm0.0095)$  \\
                          & AP  & $0.9915(\pm0.0028)$  & $0.9176(\pm0.0022)$  & $0.8461(\pm0.002)$  & $0.8077(\pm0.0062)$  & $0.8077(\pm0.0062)$  \\
                          & NE-AUC  & $\mathbf{0.9955(\pm0.0009)}$  & $0.8159(\pm0.0029)$  & $0.6972(\pm0.0087)$  & $0.6449(\pm0.0084)$  & $0.6449(\pm0.0084)$  \\
                          & NE-AP  & $0.9882(\pm0.0043)$  & $0.8588(\pm0.0025)$  & $0.7711(\pm0.0035)$  & $0.7285(\pm0.005)$  & $0.7285(\pm0.005)$  \\
  \bottomrule
  \end{tabular} %}
  \caption{Evolution pattern prediction results presented as mean values with standard deviation for both the whole graph and new edges on the cit-HepPh dataset across all models trained using $G_0$. A bold value indicates the highest score for that metric for the given graph snapshot.}
  \label{tab:HepPh}
\end{table*}

\begin{table*}[h!]
  \centering
  %\resizebox{\textwidth}{!}{
  \begin{tabular}{ll c c c c c}
  \toprule
  \textbf{Model}  & \textbf{Metric} & $G_1$           & $G_2$         & $G_3$          & $G_4$         & $G_5$ \T\B \\
  \midrule \midrule

\multirow{4}{*}{GVAE}   & AUC  & $0.9702(\pm0.0009)$  & $0.9336(\pm0.0026)$  & $0.8796(\pm0.0018)$  & $0.8503(\pm0.0018)$  & $0.8499(\pm0.0009)$  \\
                        & AP  & $0.9759(\pm0.0006)$  & $0.9461(\pm0.0021)$  & $0.9016(\pm0.0012)$  & $0.8762(\pm0.0013)$  & $0.8761(\pm0.0007)$  \\
                        & NE-AUC  & $0.9586(\pm0.002)$  & $0.9149(\pm0.0005)$  & $0.8552(\pm0.0029)$  & $0.8232(\pm0.002)$  & $0.8231(\pm0.0039)$  \\
                        & NE-AP  & $0.9521(\pm0.0016)$  & $0.9103(\pm0.0012)$  & $0.8581(\pm0.0028)$  & $0.8289(\pm0.0023)$  & $0.8289(\pm0.0034)$  \\
\midrule
\multirow{4}{*}{GAE}  & AUC  & $0.9885(\pm0.0006)$  & $0.9794(\pm0.0018)$  & $0.9545(\pm0.0011)$  & $0.9412(\pm0.0007)$  & $0.9412(\pm0.0007)$  \\
                      & AP  & $0.9902(\pm0.0004)$  & $0.9815(\pm0.0014)$  & $0.9616(\pm0.0011)$  & $0.9518(\pm0.0007)$  & $0.9518(\pm0.0007)$  \\
                      & NE-AUC  & $0.9837(\pm0.0008)$  & $0.9732(\pm0.0022)$  & $0.9447(\pm0.0015)$  & $0.9303(\pm0.0007)$  & $0.9303(\pm0.0007)$  \\
                      & NE-AP  & $0.9803(\pm0.0008)$  & $0.9685(\pm0.002)$  & $0.9442(\pm0.002)$  & $0.9332(\pm0.0012)$  & $0.9332(\pm0.0012)$  \\

\midrule
\midrule
\multirow{4}{*}{TO-GVAE}  & AUC  & $0.9957(\pm0.0004)$  & $0.9871(\pm0.0006)$  & $0.9651(\pm0.0014)$  & $0.9534(\pm0.0009)$  & $0.9524(\pm0.001)$  \\
                          & AP  & $\mathbf{0.9944(\pm0.0007)}$  & $0.9869(\pm0.0005)$  & $0.9693(\pm0.0009)$  & $0.9596(\pm0.0008)$  & $0.9591(\pm0.0011)$  \\
                          & NE-AUC  & $0.9964(\pm0.0003)$  & $0.9841(\pm0.0005)$  & $0.9578(\pm0.0022)$  & $0.9439(\pm0.0011)$  & $0.9427(\pm0.0003)$  \\
                          & NE-AP  & $0.9921(\pm0.0008)$  & $0.979(\pm0.0006)$  & $0.9549(\pm0.0021)$  & $0.9427(\pm0.0017)$  & $0.9419(\pm0.0015)$  \\
\midrule

\multirow{4}{*}{TO-GAE}   & AUC  & $\mathbf{0.9961(\pm0.0001})$  & $\mathbf{0.99(\pm0.0001)}$  & $\mathbf{0.9751(\pm0.0009)}$  & $\mathbf{0.9645(\pm0.0003)}$  & $\mathbf{0.9645(\pm0.0003)}$  \\
                      & AP  & $0.9943(\pm0.0003)$  & $\mathbf{0.9887(\pm0.0001)}$  & $\mathbf{0.9757(\pm0.0008)}$  & $\mathbf{0.967(\pm0.0008)}$  & $\mathbf{0.967(\pm0.0008)}$  \\
                      & NE-AUC  & $\mathbf{0.997(\pm0.0001)}$  & $\mathbf{0.9882(\pm0.0002)}$  & $\mathbf{0.9701(\pm0.0011)}$  & $\mathbf{0.9579(\pm0.0003)}$  & $\mathbf{0.9579(\pm0.0003)}$  \\
                      & NE-AP  & $\mathbf{0.9922(\pm0.0006)}$  & $\mathbf{0.9825(\pm0.0)}$  & $\mathbf{0.9646(\pm0.0014)}$  & $\mathbf{0.9536(\pm0.0013)}$  & $\mathbf{0.9536(\pm0.0013)}$  \\
  \bottomrule
  \end{tabular} %}
  \caption{Future link prediction results presented as mean values with standard deviation for both the whole graph and new edges on the cit-HepPh dataset across all models trained using $G_0$. A bold value indicates the highest score for that metric for the given graph snapshot.}
  \label{tab:HepPh-link}
  \vskip -20pt
\end{table*}

\subsubsection{Future Link Prediction}

Table \ref{tab:HepPh-link} highlights the results for all models at the task of future link prediction on the cit-HepPh dataset. We can see that compared with the previous task, all models are more closely matched. The results show all approaches to have a good predictive performance, even at later time points, however, the temporally offset approaches still outperform the baselines. Here it is interesting to note that the non-probabilistic model almost always out performs the variational one. 

%We hypothesise that the discrepancy in performance between the two tasks on this dataset, could indicate that 

\section{Conclusion}
\label{sec:conclusion}

Whilst a lot of focus has recently been placed on finding methods for learning graph representations which are accurately able to make predictions about the current state of the graph, few works have investigated how these models perform for temporal graphs. This paper has explored the use of graph specific neural networks to learn vertex level representations which are accurately able to make predictions about future time points of evolving graphs. This is achieved by explicitly  training the models to predict future time steps via the temporal offset reconstruction method. To assess our approach, we perform extensive experimentation on both real-world and synthetic evolving graph datasets. 

There is large scope for future continuation of this research. We plan to investigate if the use of topological vertex properties can be used as input to the model to improve performance, instead of relying upon the identity matrix of the graph. We also plan to investigate if an inductive style model \cite{hamilton2017inductive} could be used to replace the GCN as this would allow us to vary the number of vertices between time steps and allow for batch-based training. 

\section*{Acknowledgements}

We gratefully acknowledge the support of NVIDIA Corporation with the donation of the GPU used for this research. Additionally we thank the Engineering and Physical Sciences Research Council UK (EPSRC) for funding.

% references section

\bibliographystyle{IEEEtran}
\bibliography{paper_ref}

% that's all folks
\end{document}